# Defining Ideal Phantom Polymer Networks

Corresponding author/ First Author: Hemant Nanavati

Professor, Department of Chemical Engineering, Indian Institute of Technology Bombay, Mumbai – 400 076, Maharashtra, India

*E-mail: hnanavati@iitb.ac.in, hnanavati@che.iitb.ac.in

ORCID: 0000-0002-5982-6531

Second Author: Sushanta Das

Additional Director, Defence Research and Development Organization, New Delhi – 110011, India

ORCID: 0009-0006-6215-3573

## Abstract

Elastomers are modelled as networks with $\phi$-functional junctions of $N$ ideal, $n$-segment, freely jointed chains (FJCs), per unit volume (p.u.v.). Our compact model of the exact FJC length probability density (Treloar, 1975), accurately yields their exact distribution moments (Flory, 1969). The governing geometry of fluctuations of $N_X$=2$N/\phi$ junctions p.u.v., parametrically maps their $\lambda$(elongation ratio)-dependent distribution to an FJC of $n_{f\phi}$=($n/\phi$)(1–$\Lambda$) segments; $\Lambda$=(1/(3$n$))($\lambda^2$+(2/$\lambda$)). The resulting elastic pre-factor $N_{\mathrm{eff}}$k$T$=($N$– $\eta N_X$)k$T$, with junction effectiveness, $\eta$= $\phi$(1–$\Lambda$)/($\phi$–$\Lambda$), defines ideal phantom networks.

# 1 Introduction

Elastomers are three-dimensional molecular networks of chemically connected polymer chains. An ideal network is monodisperse, with each chain modelled as a finite freely jointed chain (FJC) of *n* segments of length *l*, which can freely orient and also pass through other network chain segments [1–6]. Thus, the network junctions exhibit free but finite fluctuations that naturally decrease with deformation, as quantified by our compact and accurate ideal FJC length-distribution model. The consequent deformation-dependent junction-fluctuation effectiveness defines ideal phantom elastomer networks via their resulting uniaxial stress-deformation relationships. These ideal phantom networks are the appropriate platform on which to construct real network models, which include constraint and entanglement effects [7–20].

# 2 Ideal Chains and Chain Junctions

## 2.1 Ideal FJC: Force-Extension and Length Distribution

Uncoiling increases an ideal FJC's end-to-end (vector) length, *h*, reducing its spatial pathway options, whose metric is the vector probability density, $\omega_n\left(h\right)$. In dimensionless form, $\Omega_n\left(H\right)=\left(nl\right)^3\omega_n\left(h\right)$ and $H=h/\left(nl\right)$; entropy, $s\left(h\right)/\mathrm{k}=S\left(H\right)=-\left(\ln\Omega_n+const.\right)$; k is the Boltzmann constant. For force, f, at temperature *T*, $F=\left(\mathrm{f}\,l\right)/\left(\mathrm{k}T\right)$. Work, $W=n\int_0^H F\,dH=w\left(h\right)/\left(\mathrm{k}T\right)$, is entropic [1,6]:

$$\begin{aligned} W\left(H\right) &= -\Delta S\left(H\right)=-\left(S\left(H\right)-S\big|_{H=0}\right) \\ &= -\left(\ln\Omega_n-\ln\Omega_n\big|_{H=0}\right) \end{aligned} \tag{1}$$

Exact FJC distribution functions, $\omega_n\left(h\right)=\left(1/2\pi^2h\right)\left(\int_0^\infty \sin\left(qh\right)\left(\sin\left(ql\right)/ql\right)^n q\,dq\right)$, by Rayleigh [4] and Chandrasekhar [3], are computable for small *n*. Treloar's exact alternative [1,2] is,

$$\Omega=\left(\frac{n^3\left(n-1\right)}{8\pi H}\right)\sum_{k=0}^{\mathrm{Floor}\left[\frac{1}{2}\left(n-nH\right)\right]}\frac{\left(-1\right)^k\left(\frac{1}{2}\left(-Hn-2k+n\right)\right)^{n-2}}{k!\left(n-k\right)!} \tag{2}$$

Subscript *n* is implicit. $\Omega=\Omega_0\exp\left(-W\right)$; $\Omega_0=\Omega\big|_{H=0}$ (eqn. (1)). The L'Hôpital rule yields $\Omega_0$.

$$\Omega_0=\left(\frac{n^3\left(n-1\right)}{8\pi}\right)\left(\frac{d}{dH}\sum_{k=0}^{\mathrm{Floor}\left[\frac{1}{2}\left(n-nH\right)\right]}\frac{\left(-1\right)^k\left(\frac{1}{2}\left(-Hn-2k+n\right)\right)^{n-2}}{k!\left(n-k\right)!}\right)\Bigg|_{H=0} \tag{3}$$

The limiting, small $H$, large $n$, ideal FJC, is approximated as an infinitely extensible Gaussian chain [2]; $\Omega_{\mathrm{G}} = \Omega_{0\mathrm{G}} \exp\left(-3nH^2/2\right)$; $\Omega_{0\mathrm{G}} = \left(3n/(2\pi)\right)^{3/2}$. For $n \geq 12$ (Supplementary (S.1.1))

$$\Omega_0 = \Omega_{0\mathrm{G}} / \left(1 + (0.75/n)\exp(0.5/n)\right) \tag{4}$$

To date, statistical mechanical treatments have yielded $W(H)$, via inversion of the Langevin function $\left(L(F) = \coth(F) - (1/F) = \langle H \rangle\right)$, by assuming $F = L^{-1}(H)$, because the correct, $F = L^{-1}\left(\langle H \rangle\right)$, is unusable. $L^{-1}(H)$ frameworks [21–25] based on the Padé method [26], approximate $F(H)$ [27,28] for ideal FJCs. We follow Slater [29], who introduced $1/n$ terms, to describe an ideal FJC. Generalizing Slater's expression (Supplementary (S.1.1)),

$$F = H\left(3\left(1 - (A/n)\right) - \left(1 + (B/n)\right)H^2\right) / \left(1 - H^2\right) \tag{5}$$

The odd-order polynomial numerator, has parameters $A$ and $B$; the denominator makes $F$ diverge, for the fully stretched chain $(H \to 1)$. $W(H)$ is then a compact, rotationally invariant, even function [30].

$$W = \left(1/2\right)\left((B+n)H^2 + (3A + B - 2n)\ln\left(1 - H^2\right)\right) \tag{6}$$

Fitting $\ln W$ (eqns. (1) and (6)) to numerical computations of equations (2) and (3), minimizes relative errors (Figure 1a), and yields $A \sim 1$, $B \sim 0.5n^{0.75}$. Equation (7), for our distribution's moments, $\langle H^{2t} \rangle$ ($t$=1–5), accurately yields the exact distribution moments [2] (Figure 1b) (Supplementary (S.1.1)).

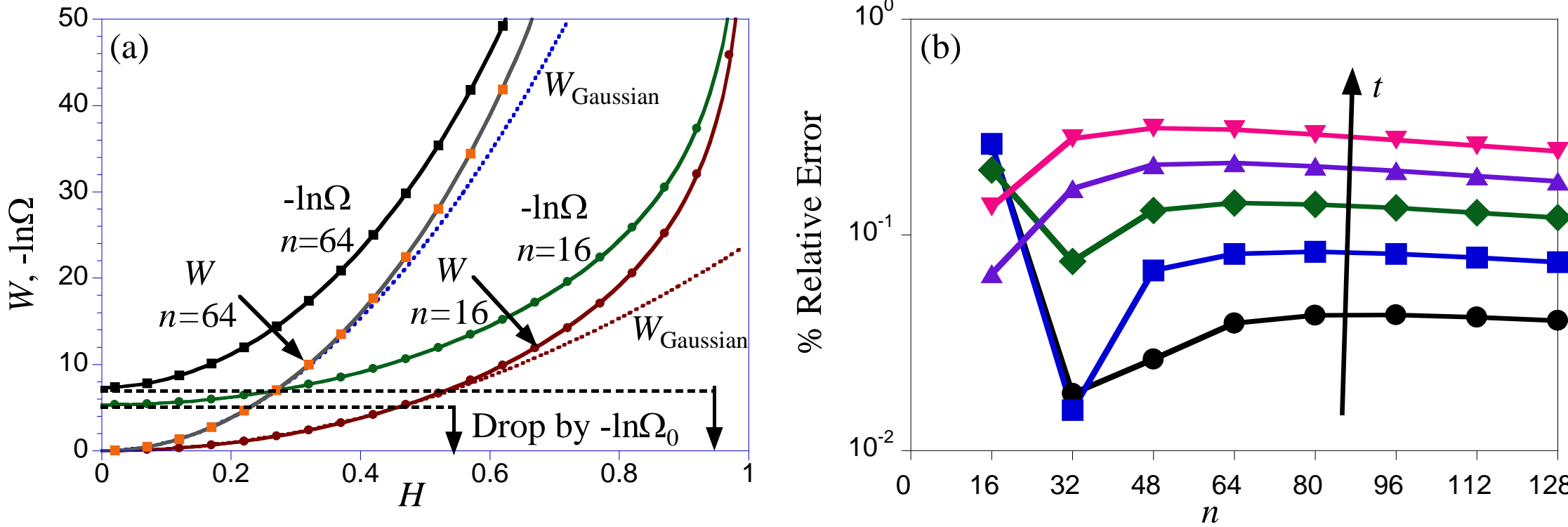

**Figure 1.** Accuracy of the compact model of exact FJCs (a) Relating $W$ to $-\ln\Omega$ via $\ln\Omega_0$ for $n$=16 and $n$=64. The curves (eqn. (6)) very closely fit the computed exact $W$ (eqn. (1)). The dotted curves describe Gaussian chains, $W_{\mathrm{G}} = 3nH^2/2$. (b) Very low relative error magnitudes (0.02% to 0.3%) for $\langle H^{2t} \rangle$ ($t$=1 to 5) from eqn. (7); $n \geq 16$. Lines aid the eye.

$$\left\langle H^{2t}\right\rangle = \int_0^1 4\pi H^2 \Omega H^{2t} dH = \int_0^1 H^{2t} P\left(H\right) dH \qquad (7)$$

$P(H)$ is the scalar distribution function. Next, our accurate and validated $W$ and $\Omega$ expressions, quantify ideal FJC junction fluctuations of phantom networks.

## 2.2 Network Junction Fluctuations

Phantom network junctions have been assumed to exhibit deformation-invariant Gaussian fluctuations [7–20,31]. However, the governing geometry of network junction fluctuations, dictates that, being finite, they should naturally decrease, as the FJCs' far ends move apart.

In Figure 2, the mean position of P (the mid-point along the *contour* AB of a $2n$-segment FJC) is at O, the mid-point of *vector* AB$\left(=i_c\right)$ ($l$=1). We consider APB, the $2n$-segment FJC, combining two $n$-segment FJCs, originating at A (vector length, AP=$h_1 = i_1 l = i_1$) and at B (BP=$h_2 = i_2$), and meeting at junction P [14]. Then, fluctuation length, OP=$i_f$; $\angle \mathrm{BOP} = \theta_f$. The fluctuation vector length distribution, $\omega_f\left(i_f\right)$ combines $\omega_n\left(i_1\right)$ and $\omega_n\left(i_2\right)$, on the condition, $\omega_{2n}\left(i_c\right)$ $\left(0 \le i_c \le 2n\right)$.

$$\omega_f\left(i_f\right) = \omega_n\left(i_1\right)\omega_n\left(i_2\right)/\omega_{2n}\left(i_c\right) \qquad (8)$$

$$i_1^2 = i_f^2 \sin^2\theta_f + \left(i_f \cos\theta_f + \left(i_c/2\right)\right)^2 ;\; i_2^2 = i_f^2 \sin^2\theta_f + \left(\left(i_c/2\right) - i_f \cos\theta_f\right)^2 \qquad (9)$$

$i_f$ ($\theta_f$) varies from 0 (P lies on O, $i_1 = i_2 = i_c/2$), to $i_{fm}$=OQ ($i_1 = n$, $i_2 = i_2^0$) (Supplementary (S.1.2.1)).

$$i_{fm} = \tfrac{1}{2}\left(-i_c \cos\theta_f + \sqrt{\left(i_c \cos\theta_f\right)^2 + 4\times\left[n^2 - \left(i_c/2\right)^2\right]}\right) \qquad (10)$$

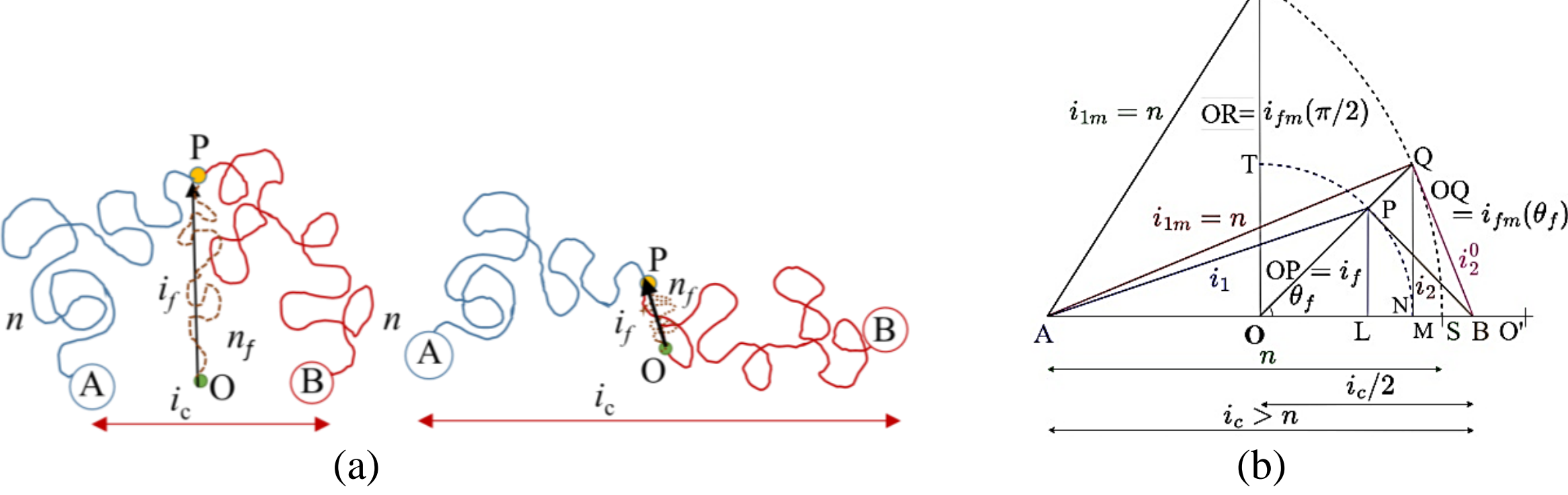

**Figure 2.** Governing geometry of junction fluctuations: (a) A schematic of a fluctuating junction of two chains, AP and BP; the fluctuation range of OP (vector length $i_f$) decreases as $i_c$, the combined AB chain end separation increases. (b) Junction fluctuation geometry for $i_c > n$; chain vectors, AP=$i_1$, BP=$i_2$; fluctuation $i_f$ =OP, its full range, at angle $\theta_f$, $i_{fm}$=OQ

$\omega_f(i_f)$ varies with $\cos\theta_f$ $(0 \le \theta_f \le \pi/2)$. The two-chain junction fluctuation $\omega_f$, parametrically maps to the distribution, $\omega(h)$, of an ideal FJC of $n_{f2}$ segments, via (i) $\Omega_{f0} = (n_{f2})^3 \omega_f|_{(i_f=0)}$ (eqn.(8) )$=\Omega_0$ (eqn.(4)) and (ii) $n_{f2} = \langle i_f^2 \rangle$, via numerical integration (eqn. (11)).

$$\langle i_f^2 \rangle = \int_0^1 \int_0^{i_{fm}(\cos\theta_f)} i_f^2 \omega_f 4\pi i_f^2 di_f d(\cos\theta_f) \tag{11}$$

Via both parameters, $\omega_{f0}$ and $\langle i_f^2 \rangle$, (i) at $i_c$=0, $n_{f2} = n/2 = n_{f2G}$; $n_{f2G}$ maps to the fluctuation of a junction of two Gaussian chains [14] and (ii) normalized $\overline{n_f} = n_{f2}/n_{f2G} = 1 - H_c^2$ (Figure 3). The fluctuation range, however, extends beyond an FJC range (Supplementary (S.1.2.2)); i.e., $i_{fm} \ge n_{f2}$.

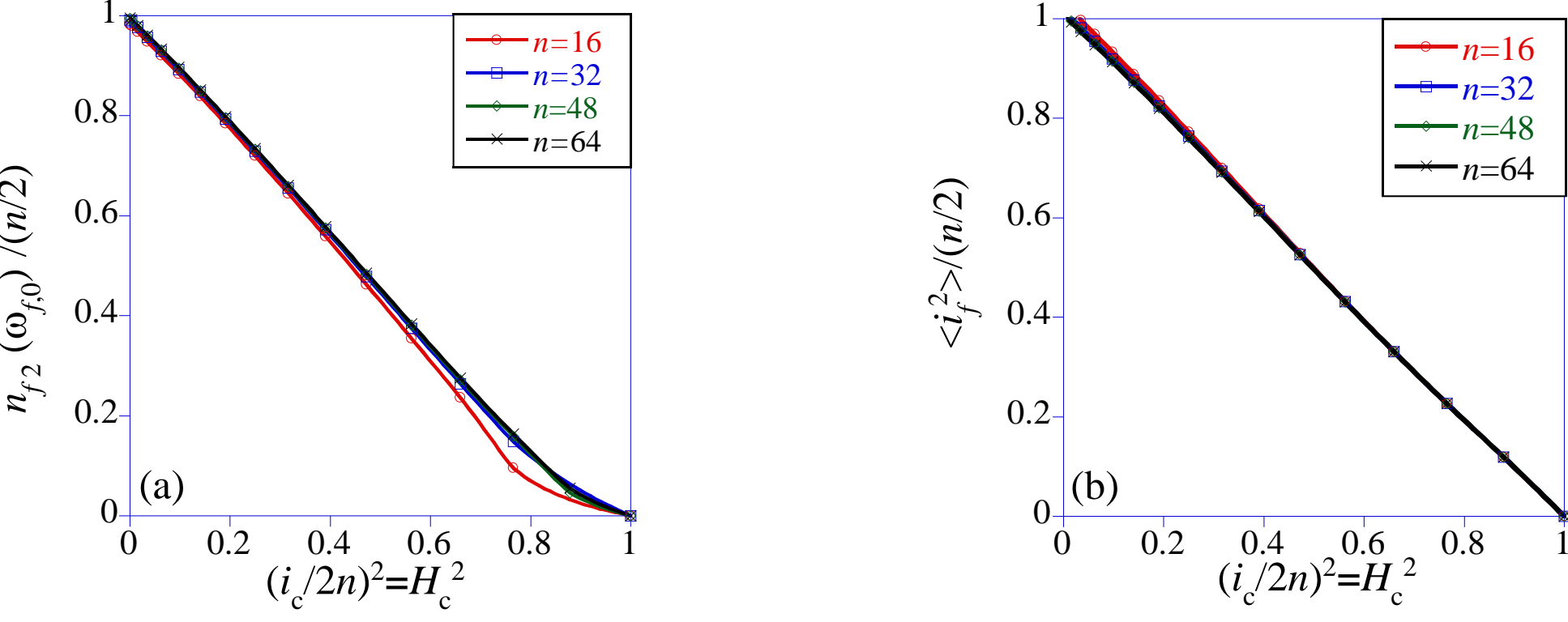

**Figure 3.** Normalized fluctuations, $\overline{n_f} = n_{f2}/(n/2) = 1 - H_c^2$; $H_c^2 = (i_c/(2n))^2$. Based on (a) $\omega_{f0}$ and (b) $n_{f2} = \langle i_f^2 \rangle$. Lines are to guide the eye.

$\langle i_f^2 \rangle$ and $n_{f2}(\Omega_{f0})$ of a 1-D two-chain junction (Supplementary (S.1.2.3)), vary in the same way as those of a 3-D two-chain junction. $\langle i_f^2 \rangle$ and $n_{f4}(\Omega_{f,0})$ behave similarly for 1-D four-chain junctions (functionality, $\phi = 4$) (Supplementary (S.1.2.4)). Then $\overline{n_f} = n_{f\phi}/(n/\phi) = n_{f\phi}/n_{f\phi G} = 1 - H_c^2$, for any $\phi$, reflects in the resulting $\sigma - \lambda$ (uniaxial stress-deformation) expressions for these ideal monodisperse 3-D phantom networks.

## 3 Affine and Phantom Networks

Ideal phantom networks are defined via (i) combining incompressible affine ideal networks with non-fluctuating junctions [1,32–40], (Supplementary (S.2.1.1)) with (ii) ideal junction fluctuations over the network; these combine Gaussian junction fluctuations [14], with finite extensibility (section 2.2).

### 3.1 Stress-Elongation of Affine Network (No Junction Fluctuations)

Network deformation comprises chain vector re-orientation and deformation. These depend on $h_0^2$, the undeformed squared vector length, and on $\Theta_0$, the undeformed chain's angle with the deformation axis [32] (Supplementary (S.2.1.1)). Affine $\lambda$ is the bulk and molecular level uniaxial deformation ratio of the deformed to the undeformed length. Uniform undeformed, $H_0^2 = h_0^2/(nl)^2 = (1/n)$; after incompressible deformation, $h^2 = h_0^2 \times \left(\lambda^2 \cos^2\Theta_0 + \left(\sin^2\Theta_0\right)/\lambda\right)$. Uniform $\cos^2\Theta_0 = \left\langle \cos^2\Theta_0 \right\rangle = 1/3$ represents all $\Theta_0$. Such undeformed chain vectors satisfy 8-chain [36] or 4-chain [34,37] models (Figure 4). These features define ideal affine networks, with non-fluctuating junctions (see section 3.3). Uniform representative deformed vector lengths, $h^2 = \left(h_0^2/3\right)\times\left(\lambda^2 + (2/\lambda)\right)$ [34,36,37]. If $\lambda|_{H\to 1} = \lambda_{\mathrm{m}}$, $\Lambda = \left(\lambda^2 + (2/\lambda)\right)/\left(\lambda_{\mathrm{m}}^2 + (2/\lambda_{\mathrm{m}})\right) = \left(1/(3n)\right)\left(\lambda^2 + (2/\lambda)\right)$. $W_{def}(\Lambda) = W - W|_{H_0}$, for the deformation of a single network chain.

$$W_{def} = \frac{1}{2}\left(\left(n + 0.5n^{0.75}\right)\left(\Lambda - \frac{1}{n}\right) + \left(3 + 0.5n^{0.75} - 2n\right)\ln\left(\frac{1-\Lambda}{1-(1/n)}\right)\right) \tag{12}$$

For $N$ monodisperse network chains (between crosslinks) per unit volume, the affine network strain energy density, $\mathbf{w}_{Af} = N\mathrm{k}TW_{def}$ and stress, $\sigma_{Af} = d\mathbf{w}_{Af}/d\lambda$; $d\Lambda/d\lambda = \left(2/(3n)\right)\left(\lambda - \left(1/\lambda^2\right)\right)$. In an ideal network, $N\mathrm{k}T = (\rho \mathrm{R} T)/M_X$; $\rho$ is the density; R is the gas constant; $M_X = nM_S$ is the molecular weight of the $n$-segment network FJC (Supplementary S.2.1.2).

$$\sigma_{Af} = \left(2/(3n)\right)\left(\lambda - \left(1/\lambda^2\right)\right)\left(d\mathbf{w}_{Af}/d\Lambda\right) = \left(2/(3n)\right)\left(\lambda - \left(1/\lambda^2\right)\right)\left(\rho \mathrm{R} T/M_S\right)\mathbf{\Sigma}_{Af} \tag{13}$$

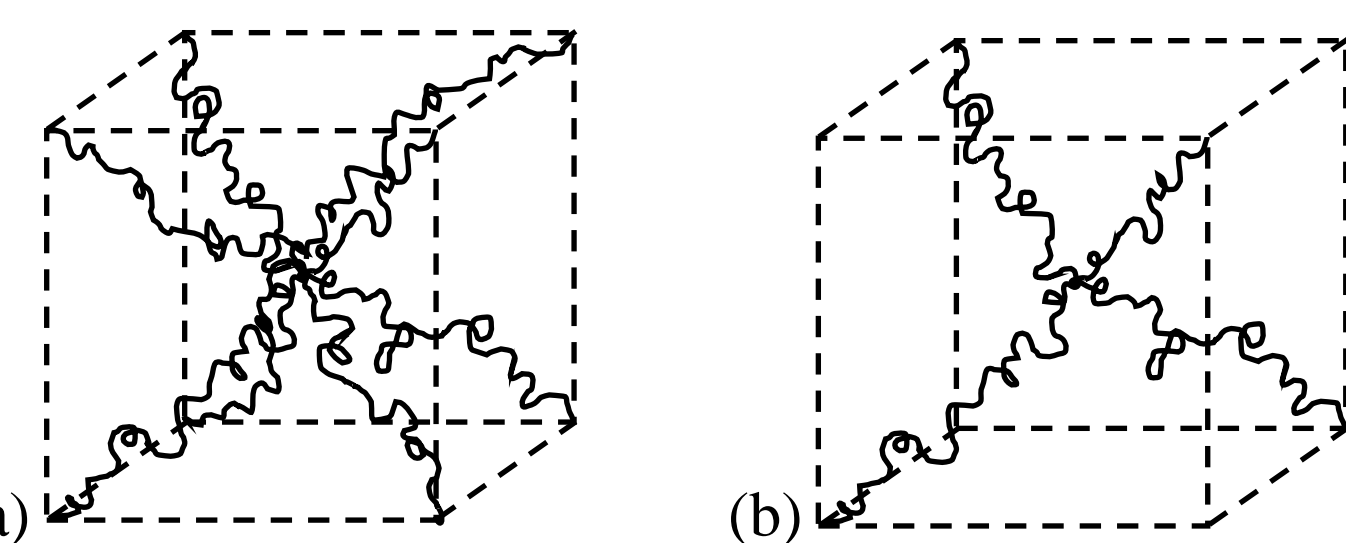

**Figure 4.** BCC-type undeformed network models: (a) 8-chain [36] and (b) 4-chain model [34,37].

Where the dimensionless form, $\mathbf{\Sigma}_{Af} = d\mathbf{W}_{Af}/d\Lambda$; $\mathbf{W}_{Af} = \mathbf{w}_{Af}/\left((\rho \mathrm{R} T)/M_S\right) = W_{def}/n$.

$$\mathbf{\Sigma}_{Af} = \left(1/(2n)\right)\left(n + 0.5n^{0.75} + \left(3 + 0.5n^{0.75} - 2n\right)/(1-\Lambda)\right) \tag{14}$$

Next, by incorporating ideal junction fluctuations, we describe ideal phantom networks.

## 3.2 Ideal Phantom Networks – Junction Fluctuation Effectiveness

*An FJC in a Network with Ideal Junction Fluctuations*

In an ideal phantom network, as the junctions fluctuate, the mean positions of the junctions move affinely (see section 3.3); i.e., $H^2 = H_c^2$. Since $H^2 = \Lambda$,

$$\overline{n_f} = 1 - \Lambda \qquad (15)$$

We then trace the path of all the fluctuations of junctions of the connected chains, from the non-fluctuating macroscopic sample boundary (the surface), up to each bulk network chain. From the bulk's surface, several groups of $\phi-1$ chains (called zero-seniority chains [14]) emanate inward, and each group meets at its first-layer junction (Figure 5). Each junction parametrically maps to an FJC sub-chain of $n_{f(\phi-1)} = n_{f(\phi-1)\mathrm{G}}\overline{n_f} = \left(n/(\phi-1)\right)\overline{n_f}$ segments, which adds to the remaining $\phi-$ functional network chain (Supplementary (S.2.2.1)). Combining the subsequent seniority network FJCs thus [14], yields $n_\infty = n\left(1+y+y^2+y^3+\ldots\right) = n/(1-y)$; $y = \overline{n_f}/(\phi-1)$. $\phi$-1 parametric FJCs of $n_\infty\overline{n_f}$ segments meet each end of the observed chain, bridging it to the surface by a sub-chain of $n_\infty y = ny/(1-y)$ segments. Then for the combined chain, $n_{\mathrm{eff}} = n + 2n_\infty y$. In an undeformed phantom network (large $n$), parametric $n_\infty y = n/(\phi-2)$, maps the fluctuations to a junction of Gaussian chains. There are no junction fluctuations at full extension; $n_\infty y = 0$. Deformation constrains intrinsic junction fluctuations progressively, and the $n_\infty y$ segments mapped from all junctions' fluctuations, revert to the junctions' fluctuation effectiveness, via $\rho \mathrm{R}T/(n_{\mathrm{eff}}M_S) = N_{\mathrm{eff}}\,\mathrm{k}T$ and $N_{\mathrm{eff}} = \left(1 - 2(1-\Lambda)/(\phi-\Lambda)\right)N$.

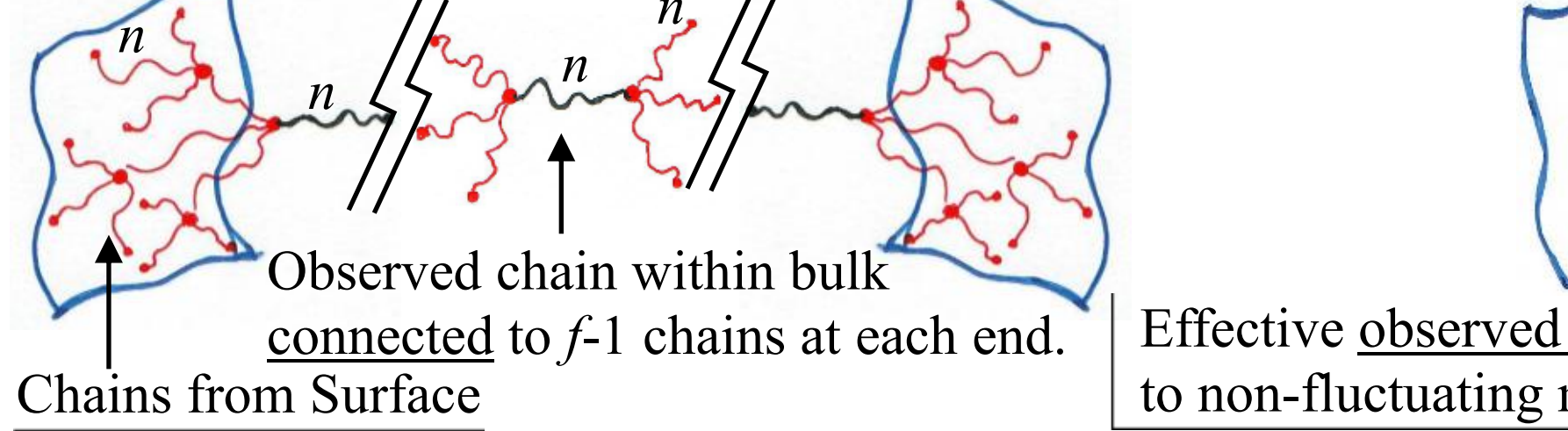

**Figure 5.** Tracing the network path from the macroscopic surfaces via connected chains and junction fluctuations, to each observed network chain. See text and Supplementary (S.2.2.1).

*Fluctuation Effectiveness of Junctions*

For $N_X = (2/\phi)N$ crosslink junctions per unit volume, the junction fluctuation effectiveness, $\eta = \phi(1-\Lambda)/(\phi-\Lambda)$; for ideal phantom networks, $\mathbf{w}_{Ph} = N_{\mathrm{eff}}\,\mathrm{k}TW_{def} = \left(N - \eta N_X\right)\mathrm{k}TW_{def}$; then $\mathbf{W}_{Ph} = \left(1 - (2\eta/\phi)\right)\mathbf{W}_{Af}$ (Supplementary (S.2.2.2)). For frameworks to date, assuming deformation-invariant fluctuation effects for Gaussian phantom [7–20,31] and finite networks [7,8,12,17,19,41], $\eta$ =1. For affine networks, $\eta$=0. Thus, ideal phantom networks deform from an undeformed Gaussian-like phantom state, into a fully stretched, affine state.

$$\mathbf{W}_{Ph} = \frac{1}{2n}\left(1-\frac{2(1-\Lambda)}{\phi-\Lambda}\right)\left(\left(0.5n^{0.75}+n\right)\left(\Lambda-(1/n)\right)-\left(2n-3-0.5n^{0.75}\right)\ln\left(\frac{1-\Lambda}{1-(1/n)}\right)\right) \quad (16)$$

$$\begin{aligned}\mathbf{\Sigma}_{Ph} &= \frac{\left(0.5n^{0.75}+n\right)}{2n}\left(\left(1-\frac{2(1-\Lambda)}{\phi-\Lambda}\right)+\left(1-\frac{(1-\Lambda)}{\phi-\Lambda}\right)\left(\frac{2}{\phi-\Lambda}\right)\left(\Lambda-(1/n)\right)\right)\\ &+\frac{\left(2n-3-0.5n^{0.75}\right)}{2n}\left(\left(1-\frac{2(1-\Lambda)}{\phi-\Lambda}\right)\frac{1}{(1-\Lambda)}-\left(1-\frac{(1-\Lambda)}{\phi-\Lambda}\right)\frac{2}{(\phi-\Lambda)}\ln\left(\frac{1-\Lambda}{1-(1/n)}\right)\right)\end{aligned} \quad (17)$$

In an illustrative example ($n$=20, $\phi$=4), tensile $\sigma-\lambda$ plots of finite phantom and affine networks are compared in Figure 6a. They are understood in Figure 6b, where $\sigma_{Ph}/\sigma_{Af}$ and the corresponding $\mathbf{W}_{Ph}/\mathbf{W}_{Af}$ ratios are plotted vs any uniaxial Λ. Undeformed ($\Lambda=1/n$) $\sigma_{Ph}/\sigma_{Af}=\mathbf{W}_{Ph}/\mathbf{W}_{Af}\sim 0.52$, while $\sigma_{Ph}=\sigma_{Af}=0$. The ratios become ~0.5 for $\Lambda\to 0$ (large $n$). $\Lambda<1/n$ is unphysical. Both, $\sigma_{Ph}$ and $\mathbf{W}_{Ph}$, stiffen with Λ. Then, over ~0.64<Λ<1, $\sigma_{Ph}>\sigma_{Af}$, and $\sigma_{Ph}/\sigma_{Af}$ exhibits a maximum.

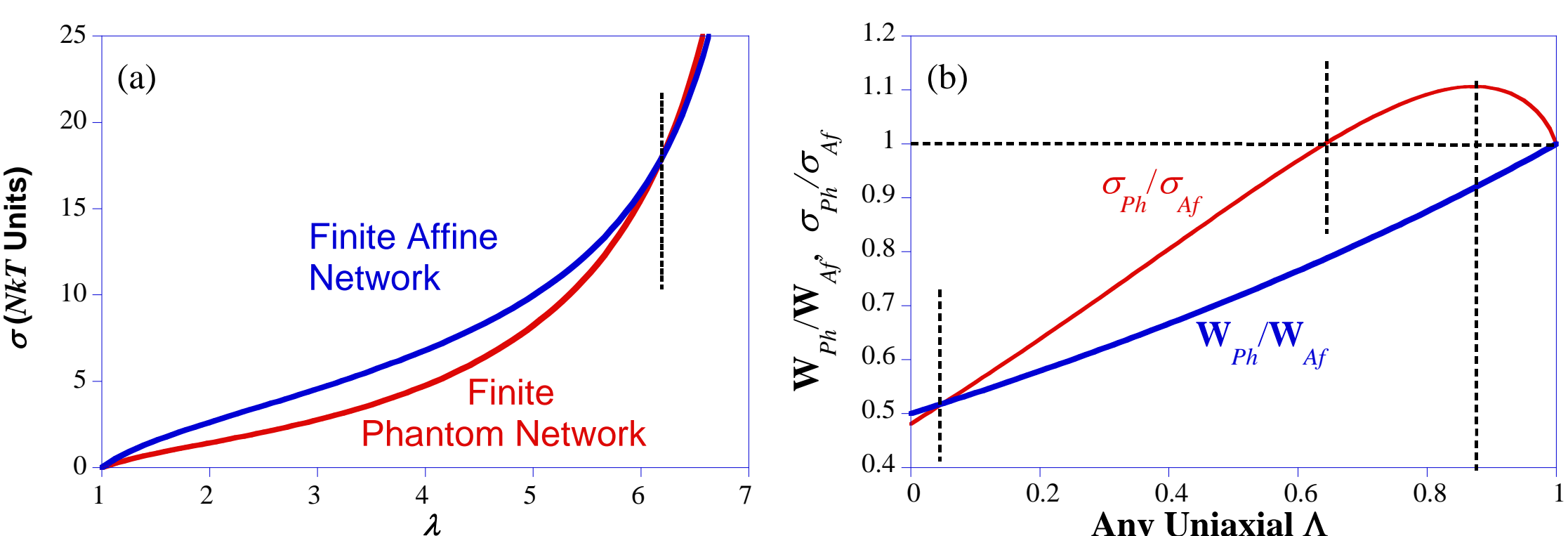

**Figure 6.** Comparing ideal phantom and affine networks ($n$=20, $\phi$=4): (a) tensile $\sigma-\lambda$ plot for phantom networks and affine networks, (b) $\sigma_{Ph}/\sigma_{Af}$ and $\mathbf{W}_{Ph}/\mathbf{W}_{Af}$ vs any uniaxial Λ. **W** varies from phantom network to affine network behaviour; over ~0.64<Λ<1, $\sigma_{Ph}/\sigma_{Af}$ >1.

Qualitatively simplified equations based on Cohen's Padé approximation [26] and linearized $\eta$, enable visualization of the mathematical skeleton of this framework. (Supplementary (S.2.2.2))

$\mathbf{W}_{Af}\sim\left(\Lambda-\frac{1}{n}\right)+2\ln\left(\frac{1-\Lambda}{1-(1/n)}\right)$; then $\mathbf{\Sigma}_{Af}\sim 1+\frac{2}{1-\Lambda}$. If $\eta\sim(1-\Lambda)$, $\frac{\mathbf{W}_{Ph}}{\mathbf{W}_{Af}}\sim\left(1-\frac{2(1-\Lambda)}{\phi}\right)$; then

$$\mathbf{\Sigma}_{Ph}\sim\left(1-\frac{2(1-\Lambda)}{\phi}\right)\left(1+\frac{2}{1-\Lambda}\right)+\frac{2}{\phi}\left(\left(\Lambda-\frac{1}{n}\right)-2\ln\left(\frac{1-\Lambda}{1-(1/n)}\right)\right).$$

### 3.3 Consequences of defining ideal phantom networks

Defining ideal networks, featuring closed-form mathematical accuracy and physical objectivity, with fundamentally self-consistent assumptions, is work in progress. In our definition of an ideal phantom network, without additional parameters, we consider (i) uniform $n$ for all network chains or monodisperse chain molecular weight, (ii) uniform segment size, $l$ (iii) uniform undeformed $H_0^2 = (1/n)$; beyond the earlier argument [33], the chain fluctuation time-scale is much shorter than the bulk deformation time-scale (consistent with the statistical mechanics-based eqn. (1)); i.e., equilibration is approached throughout deformation, precluding the reported [42–49] effects of the instantaneous undeformed vector-length stretch distribution. (iv) uniform, representative $\cos^2\Theta_0 = \langle \cos^2\Theta_0 \rangle = 1/3$. Since tensile $\lambda_{\mathrm{m}}$ increases from $\lambda_{\mathrm{m}}\big|_{\Theta_0=0} = \sqrt{n}$ to $\lambda_{\mathrm{m}}\big|_{\Theta_0=\pi/2} \to \infty$, uniform $\Theta_0$ yields uniform uniaxial $\lambda_{\mathrm{m}}$. During affine deformation from the cube-diagonal $\Theta_0$, chains only extend; i.e., chain entropy decreases uniformly and monotonically (Supplementary (S.2.1.1)). Here, ongoing efforts [19,32,34,35,48–59] via non-affine micro-macro transitions, aim to unambiguously address the effect of $\Theta_0$ distribution on chain deformation, suggesting objective criteria such as the orienting entropic force component [56] and free energy minimization [57].

Molecular models of real networks should be built on an accurate and rational ideal phantom network platform over two levels. At the intrinsic level of the chain and network, in addition to micro-macro transitions, some of the specific considerations from above, can be parametrically generalized via (i) distribution of $l$, or deformable segments [43,50,60–78] (ii) chain contour polydispersity ($n$ distribution) [42,43,50,52,79–91].

The second level is that of interchain interactions leading to surrounding constraints and entanglement effects. As seen when considering deformation-invariant intrinsic junction fluctuations [7–20], these constraints on the junction fluctuations, relax with deformation, and usually become negligible, while $\sigma_{Ph} < \sigma_{Af}$. At the other extreme are models that include chain fluctuations and their constraints, but have non-fluctuating junctions [32,51,54,92–95]. Both these include works [19,54] that consider the original Rayleigh-Chandrasekhar expression [2–4].

The ideal phantom network platform, accounts for the competing effects of decays with deformation of both, (i) intrinsic junction fluctuations and (ii) surrounding constraints (Supplementary (S.2.2.3)) to junction fluctuations. The resulting dimensionless $\boldsymbol{\Sigma}$, including both, phantom and constraint effects, scales with the Mooney-Rivlin reduced stress [1,10,14], $\sigma_{red} = \sigma / \left(\lambda - \left(1/\lambda^2\right)\right)$; it is normalized with deformation [96–99], to enable insights into the experimental sample's network structure.

# 4 Conclusions

Defining an ideal phantom network follows three stages. A compact, accurate and validated distribution function of an ideal, *n*-segments freely jointed chain (FJC), reveals that the governing geometry of the fluctuation of junctions of $\phi$ such ideal FJCs, parametrically maps the fluctuation to an ideal FJC of deformation-dependent $n_{f\phi}$ segments. $\overline{n_f} = n_{f\phi}/(n/\phi) = 1-\Lambda$ is the normalized $n_{f\phi}$; $\Lambda = (1/(3n))(\lambda^2 + (2/\lambda))$. Network deformation naturally suppresses intrinsic junction fluctuation. Tracing the network paths of fluctuating junctions of connected chains, from the macroscopic surface to each network chain, reveals the deformation-dependent effectiveness of junction fluctuations, $\eta = \phi(1-\Lambda)/(\phi-\Lambda)$. In the elasticity pre-factor, $N_{\text{eff}}\,\mathrm{k}T = (N - \eta N_X)\,\mathrm{k}T$, $\eta$ characterizes ideal phantom networks. Such phantom network expressions would be the appropriate platform on which to incorporate real network molecular models.

# Supplementary Material:

# Defining Ideal Phantom Polymer Networks

Corresponding author/ First Author: Hemant Nanavati

Professor, Department of Chemical Engineering, Indian Institute of Technology Bombay, Mumbai – 400 076, Maharashtra, India

*E-mail: hnanavati@iitb.ac.in, hnanavati@che.iitb.ac.in

ORCID: 0000-0002-5982-6531

Second Author: Sushanta Das

Additional Director, Defence Research and Development Organization, New Delhi – 110011, India

ORCID: 0009-0006-6215-3573

# S.1. Ideal FJC Chains and Networks

### S.1.1 FJC Distribution via Accurate ILF Expressions

We first describe in detail, the determination of analytical closed forms of the distribution functions of the end-to-end length (vector length), *h*, of ideal FJCs (*n* segments of length *l*) [1–6]. The vector probability distribution or the probability density, $\omega_n(h)$, is the basis of the entropic elasticity. In dimensionless form, $\Omega_n(H)=(nl)^3\,\omega_n(h)$; $H=\dfrac{h}{nl}$. $s(h)/\mathrm{k}=S(H)=-(\ln\Omega_n+const.)$ is the dimensionless entropy. k is Boltzmann's constant. f is the isothermal entropic deforming force on a chain at temperature *T*. Then, dimensionless force, $F=\left(\dfrac{\mathrm{f}\,l}{\mathrm{k}T}\right)$. $W=\dfrac{w}{\mathrm{k}T}=\displaystyle\int_0^h\frac{\mathrm{f}}{\mathrm{k}T}dh=n\int_0^H FdH$ is the dimensionless work. From statistical mechanical analyses [4–6],

$$\begin{aligned} W(H)&=-\Delta S(H)=-\left(S(H)-S|_{H=0}\right)\\ &=-\left(\ln\Omega_n-\ln\Omega_n|_{H=0}\right)\end{aligned} \tag{S.1}$$

For an ideal FJC (explicitly finite, $H\le 1$), exact distribution functions, by Rayleigh [1,2] and Chandrasekhar [2,3] (eqn. (S.2)), are computable only for small *n*.

$$\omega(h)=\frac{1}{2\pi^2 h}\int_0^\infty \sin(qh)\left(\frac{\sin(ql)}{ql}\right)^n qdq \tag{S.2}$$

Treloar's alternative exact expression [2,4], in dimensionless form, with implicit subscript *n,* is

$$\Omega=\left(\frac{n^3(n-1)}{8\pi H}\right)\sum_{m=0}^{\mathrm{Floor}\left[\frac{1}{2}(n-nH)\right]}\frac{(-1)^m\left(\frac{1}{2}(-Hn-2m+n)\right)^{n-2}}{m!(n-m)!} \tag{S.3}$$

From equation (S.1), the ideal FJC $\Omega=\Omega_0\exp(-W)$; $\Omega_0=\Omega|_{H=0}$ (eqn.(S.1)). $\Omega_0$ is obtained from equation (S.3), via the L'Hôpital rule.

$$\Omega|_{H=0}=\left(\frac{n^3(n-1)}{8\pi}\right)\left.\left(\frac{d}{dH}\sum_{m=0}^{\mathrm{Floor}\left[\frac{1}{2}(n-nH)\right]}\frac{(-1)^m\left(\frac{1}{2}(-Hn-2m+n)\right)^{n-2}}{m!(n-m)!}\right)\right|_{H=0} \tag{S.4}$$

The limiting, small $H$, large *n*, Gaussian chain approximation [2,4] of an ideal FJC, permits the vector length to extend to infinity.

$$\Omega_{\mathrm{G}} = \Omega_{0\mathrm{G}} \exp\left(-\frac{3nH^2}{2}\right) \tag{S.5}$$

We compute the numerical values of $\Omega_0$ (eqn. (S.4)) for $n \geq 12$. $\Omega_0$ relates to *n*, in terms of $\Omega_{0\mathrm{G}} = \left(3n/(2\pi)\right)^{3/2}$. Figure S.2(a) is the plot of $(\Omega_{0\mathrm{G}}/\Omega_0 - 1)$ vs 1/*n*, with a zero intercept, and an initial slope ~0.75. Figure S.2(b) contains the plot of $\ln\left((\Omega_{0\mathrm{G}}/\Omega_0 - 1)/(0.75/n)\right)$ vs 1/*n*, which is linear with slope 0.5 and passes through the origin. The resultant relationship is,

$$\Omega_0 = \frac{\Omega_{0\mathrm{G}}}{1 + \frac{0.75}{n}\exp\left(\frac{0.5}{n}\right)} \tag{S.6}$$

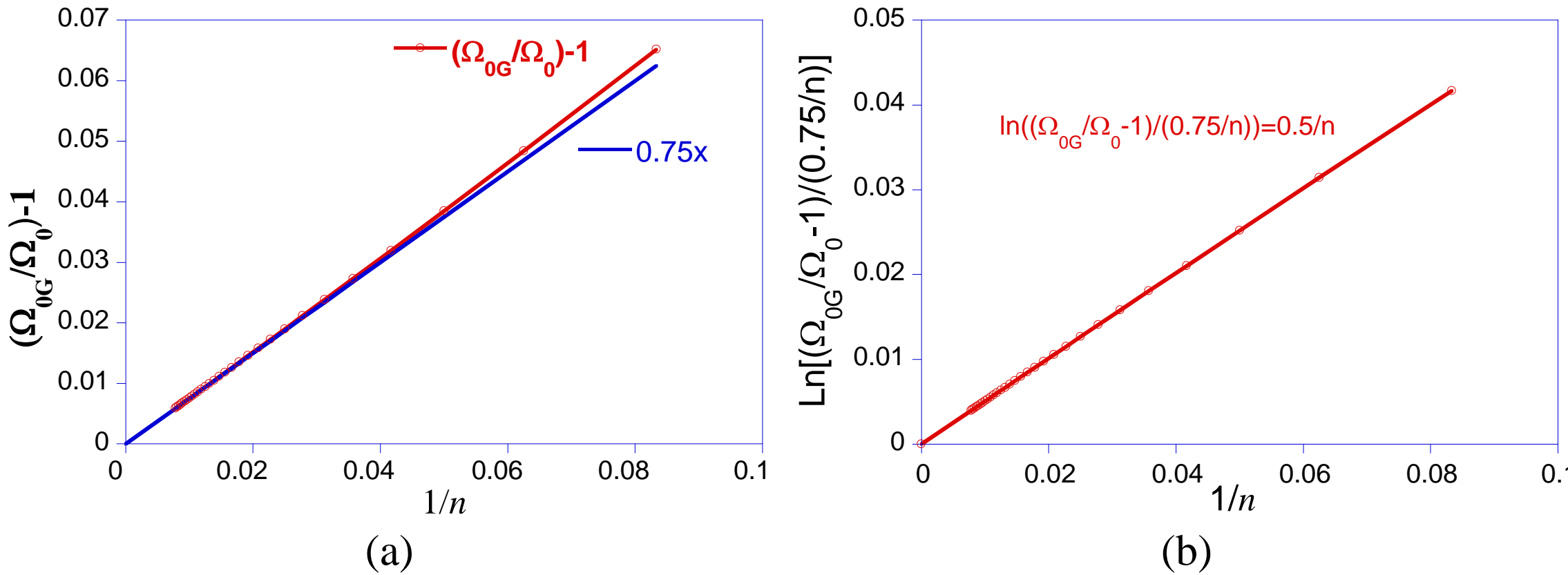

Figure S.1. (a) $(\Omega_{0\mathrm{G}}/\Omega_0 - 1)$ vs 1/*n*, with initial slope ~0.75, with a zero intercept. (a) Variation of $(\Omega_{0\mathrm{G}}/\Omega_0 - 1)$ from the initial slope of Figure S.1 (b) $\ln\left((\Omega_{0\mathrm{G}}/\Omega_0 - 1)/(0.75/n)\right)$ vs 1/*n*.

Estimates for $W(H)$ of finite chains, have been obtained via the inverse Langevin function ($L^{-1}$ or ILF) ( $L(F) = \coth(F) - (1/F) = \langle H \rangle$), assuming $F = L^{-1}(H)$, because the correct $F = L^{-1}(\langle H \rangle)$ is unusable. Cohen's [7] Padé approximation for the ILF is,

$$F = H\frac{3 - H^2}{(1 - H^2)} \tag{S.7}$$

Slater [8] has estimated an ideal FJC *F*-*H* relationship, by adding $1/n$ terms in equation (S.7).

$$F = H\frac{3\left(1 - 1/n\right) - \left(1 + 1/n\right)H^2}{(1 - H^2)} \tag{S.8}$$

Thus, the numerator is an odd-order polynomial, while the denominator causes the force to diverge for the fully stretched chain $(H \to 1)$. There is ongoing work towards developing more accurate, closed-form representations for $L^{-1}(H)$ [9–13]. Morovati and Dargazany [14] and Darabi *et al.* [15] have reported their analysis, where they incorporate *n*-dependent numerical correction terms, to the ILF, and have indicated that their maximum deviation in lnΩ as obtained by Morovati *et al.* [9], is less than 0.01. At this time, we consider ILF expressions [9,14], which are of the form,

$$L^{-1}(H) = c_0 \frac{2H}{\left(1-H^2\right)} + \sum_{i=1}^{m} c_i H^{(2i-1)} \tag{S.9}$$

The errors associated with increasing values of *m*, are 20% at *m*=0, progressively decreasing to 2.5%, 0.57%, 0.20%, 0.13% and 0.02%, at *m*=5. We generalize the method by Slater *et al.* [8] (eqn. (S.8)), to describe an ideal FJC, for *m*=1 (eqn. (S.10)) and *m*=3 (eqn. (S.11)).

$$F = H \frac{3\left(1-\left(A/n\right)\right)-\left(1+\left(B/n\right)\right)H^2}{\left(1-H^2\right)} \tag{S.10}$$

$$F = \frac{2\left(1-A/n\right)H}{\left(1-H^2\right)} + \left(1-B/n\right)H - \frac{H^3}{5} - \frac{2}{5}H^5 \tag{S.11}$$

The dimensionless force, with an odd polynomial in the numerator, yields an even function for the deformation work of a single chain [16]. For *m*=1 and *m*=3, respectively,

$$W = \left(1/2\right)\left(\left(B+n\right)H^2 + \left(3A+B-2n\right)\ln\left(1-H^2\right)\right) \tag{S.12}$$

$$W = -\left(\frac{H^2}{2}(B-n) + \frac{nH^4}{20} + \frac{H^6 n}{15} + (n-A)\ln\left(1-H^2\right)\right) \tag{S.13}$$

The simplification (Table S.1) of coefficients, $c_i$, in the ILF expressions (eqn.(S.9)), does not affect the accuracy of the ideal FJC distribution, as described next.

**Table S.1.** Coefficients considered for the Models of the ILF [9,14].

| | *m*=1 | *m*=1 (simplified) | *m*=3 | *m*=3 (simplified) |
|---|---|---|---|---|
| $c_0$ | 0.975 | 1 | 1.002 | 1 |
| $c_1$ | 0.975 | 1 | 0.9978 | 1 |
| $c_2$ | | | -0.2086 | -1/5 |
| $c_3$ | | | -0.4213 | -2/5 |

Since the values of $W$ span several orders of magnitude over the range $H$=0 to $H$=0.99, we need to minimize the relative errors in $W$. Hence, we fit (for parameters, $A$ and $B$) the expressions for $\ln W$ (via eqns. (S.12) and (S.13)) to the values of computed $\ln\left(-\left(\ln\Omega-\ln\Omega\big|_{H=0}\right)\right)$ (eqns. (S.3) and (S.4)), for various $n$ (Figure S.2). For $m$=1, $A\sim 1, B\sim 0.5n^{0.75}$; for $m$=3, $A\sim 2$, $B\sim -\left(1+\left(2/n\right)\right)$. In the simultaneous limits of large $n$ and small $H$, these FJCs are Gaussian; $W_{\mathrm{G}}=3nH^2/2$.

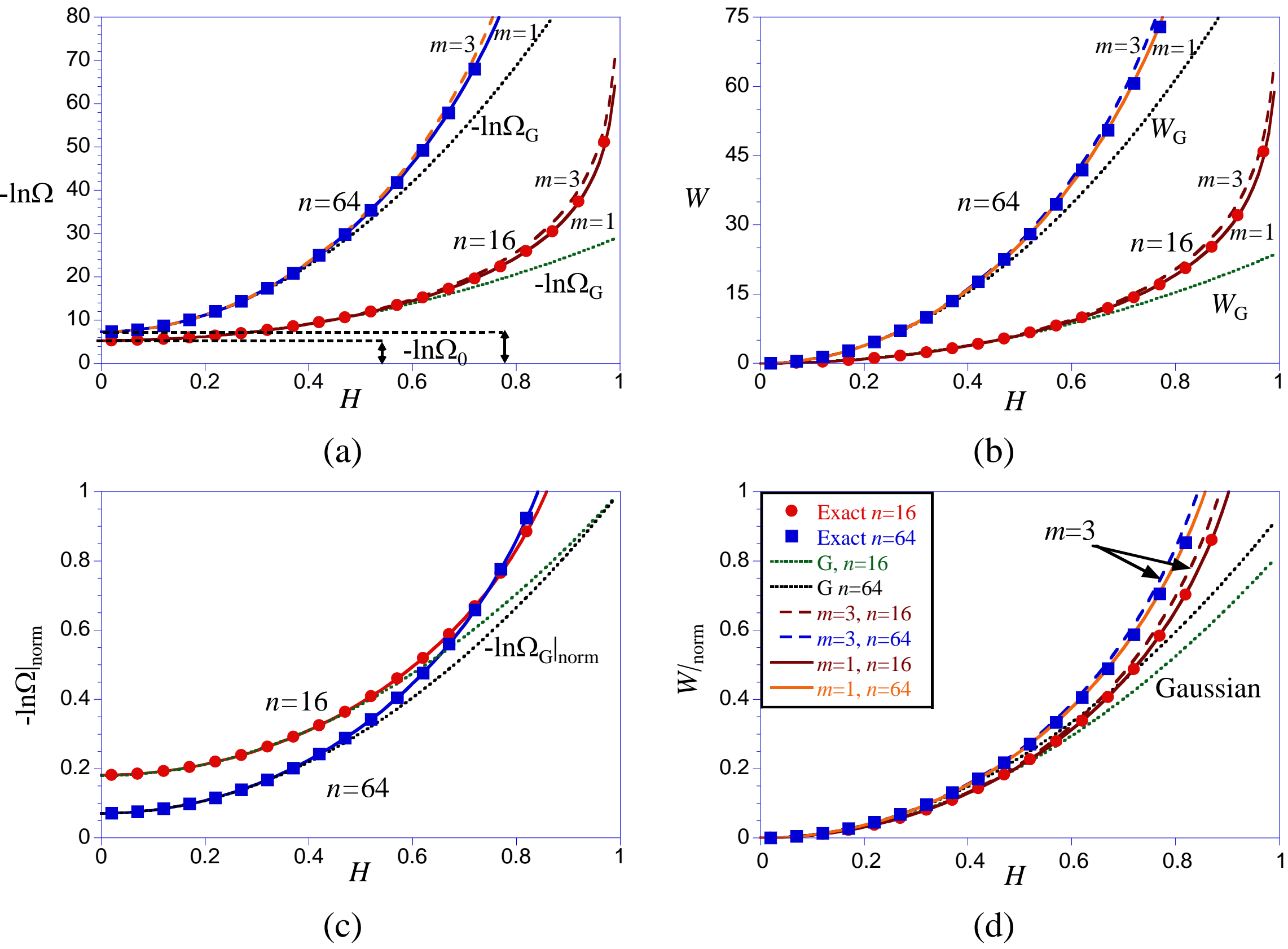

Figure S.2. $-\ln\Omega$ and $W$ plots for $n$=16 and $n$=64. The dotted curves describe Gaussian chains. In (a), (b), (d), curves are fits of the $m$=1 (eqn. (S.12)) and the $m$=3 models (eqn. (S.13)). (a) $-\ln\Omega$ plots (b) $W$ plots. (c) Normalized plots of $-\ln\Omega\Big/\left|\ln\left(\Omega_{\mathrm{G}}\big|_{H=1}\right)\right|$. Curves indicate fits of the $m$=1 model (d) Normalized plots of $W\Big/\left|\ln\left(\Omega_{\mathrm{G}}\big|_{H=1}\right)\right|$).

The relative errors for the various models are shown in Figure S.3. We verify:

$$\oint_V \omega(h)\,dV=\int_0^{na}4\pi h^2\omega(h)\,dh=\int_0^{na}P(h)\,dh=1 \tag{S.14}$$

$$\int_0^1 4\pi H^2\Omega(H)\,dH=\int_0^1 P(H)\,dH=1 \tag{S.15}$$

$P(h)$ and $P(H)$ are the scalar length probability distributions. We also compare (Figure S.4) the moments of the distribution (eqn. (S.16)), with their exact values (eqns. (S.17) to (S.21), [2]).

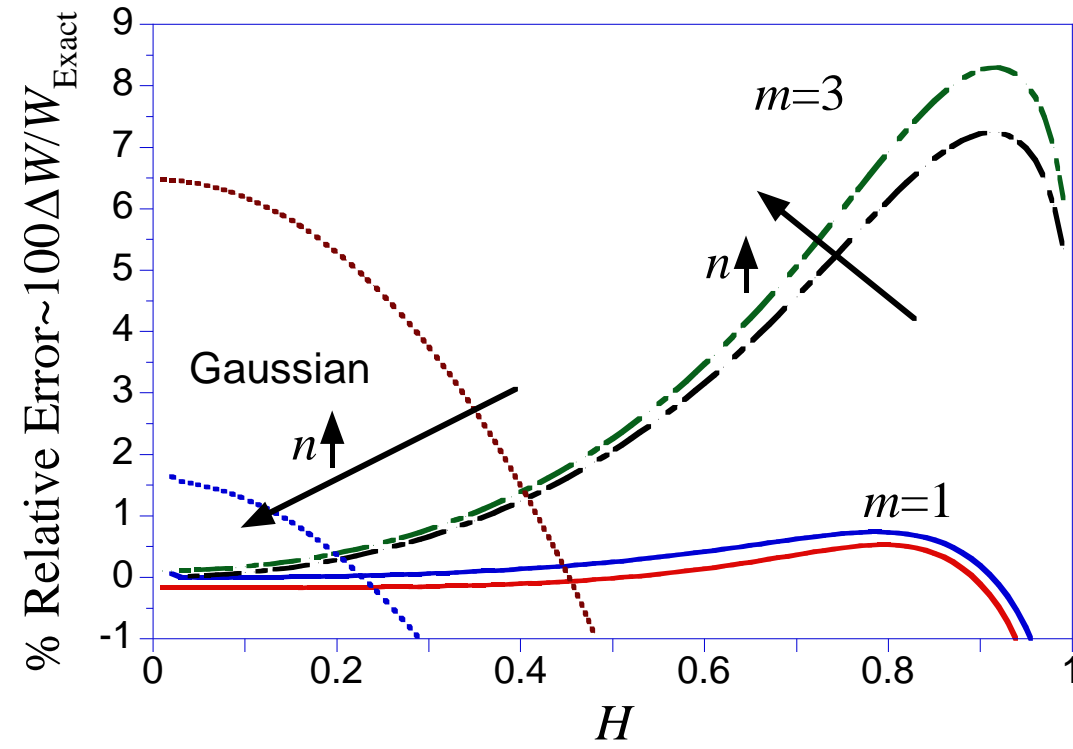

Figure S.3. % Relative errors in various $W$ Models.

$$\left\langle H^{2t}\right\rangle = \int_0^1 4\pi H^2 \Omega(H) H^{2t} dH = \int_0^1 H^{2t} P(H) dH \tag{S.16}$$

$$\left\langle H^{2}\right\rangle = 1/n \tag{S.17}$$

$$\frac{\left\langle H^{4}\right\rangle}{\left\langle H^{2}\right\rangle^{2}} = \frac{1}{n^4}\left(\frac{5}{3}n(n-1)+n\right) \tag{S.18}$$

$$\frac{\left\langle H^{6}\right\rangle}{\left\langle H^{2}\right\rangle^{3}} = \frac{1}{n^6}\left(\frac{35}{9}n(n-1)(n-2)+7n(n-1)+n\right) \tag{S.19}$$

$$\frac{\left\langle H^{8}\right\rangle}{\left\langle H^{2}\right\rangle^{4}} = \frac{1}{n^8}\left(\frac{35}{3}n(n-1)(n-2)(n-3)+42n(n-1)(n-2)+\frac{123}{5}n(n-1)+n\right) \tag{S.20}$$

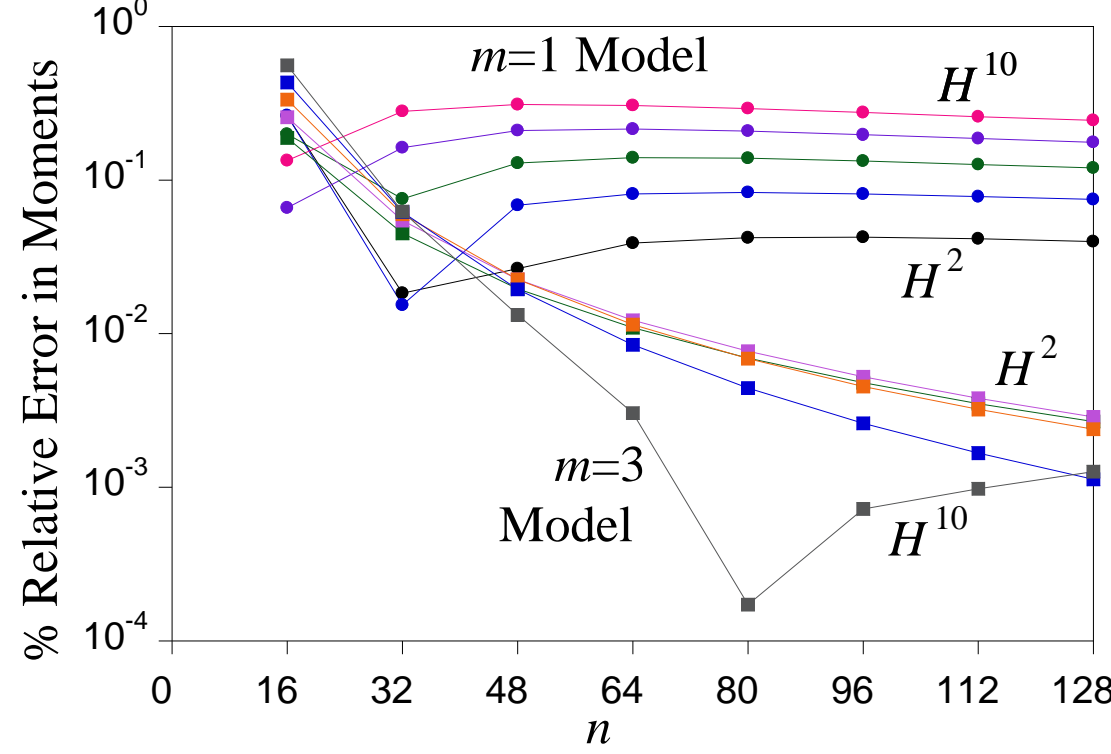

Figure S.4. Relative errors of moments <0.5% ($n \geq 16$, $m$=1 and $m$=3 models). Despite errors in $W$, (Figure S.3), errors in the moments of the distributions are very small. Lines aid the eye.

$$\frac{\langle H^{10}\rangle}{\langle H^{2}\rangle^{5}}=\frac{1}{n^{10}}\left(\begin{array}{l}\frac{385}{9}n(n-1)(n-2)(n-3)(n-4)+\frac{770}{3}n(n-1)(n-2)(n-3)\\+341n(n-1)(n-2)+\frac{253}{3}n(n-1)+n\end{array}\right) \tag{S.21}$$

Despite errors in $-\ln\Omega$ (due to the approximate fits of *A* and *B*, as functions of *n*), for $n\geq 16$, the moments, for both, the *m*=1 and *m*=3 models, accurately predict the exact FJC moments (Relative Errors < 0.5%). Integrating the simple Slater expression (eqn. (S.8)), yields errors up to ~6% (not shown). Next, our accurate models quantify the fluctuations of network junctions of ideal FJCs.

### S.1.2 Network Junction Fluctuations

This section reveals systematically, the governing geometry of junction fluctuations and their distributions (considered Gaussian to date [17–31]). We begin with the fluctuations of a two-chain junction of ideal FJCs, as function of the distance between the far ends of the junction chains.

#### S.1.2.1 Fluctuation of Two-Chain Junctions of 3-D FJCs

Figure S.5 depicts APB, the combined FJC of $n_c$ =2*n* segments of vector length, AB=$i_c$. The mean position of the *contour* mid-point P, is at O, the *vector* mid-point of AB [26]. Then, two FJCs of $n$ segments (for convenience, *l*=1), one originating from A (vector length AP=$h_1=i_1 l=i_1$) and one from B (BP=$h_2=i_2$), meet at the junction, P. OP=$i_f$, is the junction fluctuation vector length. The greatest fluctuations occur when $i_c$ =0 (A and B coincide), and become zero when $i_c$ =2*n* (fully extended AB).

We detail step-wise here, the description of the governing geometry. For clarity, we show only the chain vector representations. In Figure S.6 onward, $i_c > n$ ($\mathrm{AS}=n$). With respect to the notional +X direction (along AOB), $\angle\mathrm{BOP}=\theta_f$, $\angle\mathrm{OAP}=\theta_1$. Then $\angle\mathrm{O'BP}=\theta_2$; $\angle\mathrm{OBP}=\pi-\theta_2$. A vertical line from P intersects AB at L. We transform next, variables $i_1$ and $i_2$, to $i_f$ and $\theta_f$, given $i_c$.

$$\mathrm{LP}=i_f\sin\theta_f=i_1\sin\theta_1 \tag{S.22}$$

$$\mathrm{OL}=i_f\cos\theta_f=\mathrm{AL}-\mathrm{AO}=i_1\cos\theta_1-\left(i_c/2\right) \tag{S.23}$$

$$i_1^2=i_f^2\sin^2\theta_f+\left(i_f\cos\theta_f+\left(i_c/2\right)\right)^2 \tag{S.24}$$

$$i_2^2=i_f^2\sin^2\theta_f+\left(\left(i_c/2\right)-i_f\cos\theta_f\right)^2 \tag{S.25}$$

The ranges of $i_1$ and $i_2$, relate to the corresponding ranges of $i_f$ and $\theta_f$ (Figure S.7). $i_1 + i_2 \geq i_c$, and $i_1$ can vary from 0 to *n*. The corresponding $i_2$ varies from $|i_c - i_1|$ to *n*, as $\theta_1$ increases from 0 towards $\pi$, via cooperative variation in $\theta_2$, resulting in the variation in $i_f$ and $\theta_f$. At any $i_c$, $i_{1m} = n$ is the maximum value of $i_1$. Given $i_c$, at any $\theta_f$, $i_f$ varies from 0 (P lies on O, $\theta_1$=0 and $i_1 = i_2 = i_c/2$), to $i_f = i_{fm}$ (fluctuation length = OQ, $i_1 = i_{1m} = n$, $\theta_1 = \theta_1^0$). This increase in $i_f$ is associated with a concerted increase in $i_1$ and $\theta_1$. Then, equations (S.22) to (S.25), can be re-framed for OQ, OM, $i_{1m}$ and $i_2^0$. $i_2^0$ is not the maximum value of $i_2$, but the value of $i_2$, for a given $\theta_f$, corresponding to $\theta_1 = \theta_1^0$ (with $\theta_2 = \theta_2^0$) and $i_1 = i_{1m} = n$. The relationship between $i_{fm}$ and $\theta_f$, is obtained by replacing in equation (S.24), $i_f$ with $i_{fm}$, and $i_1$ with $i_{1m} = n$. Then,

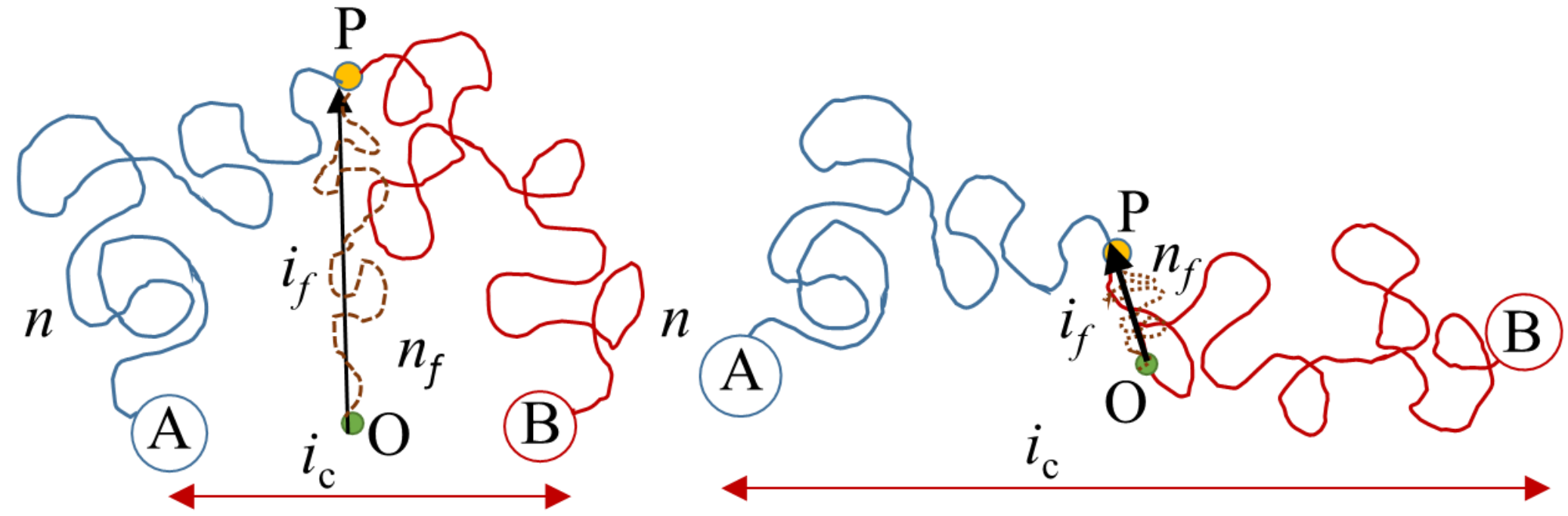

Figure S.5. A Schematic of a fluctuating junction of two chains. Description in the text.

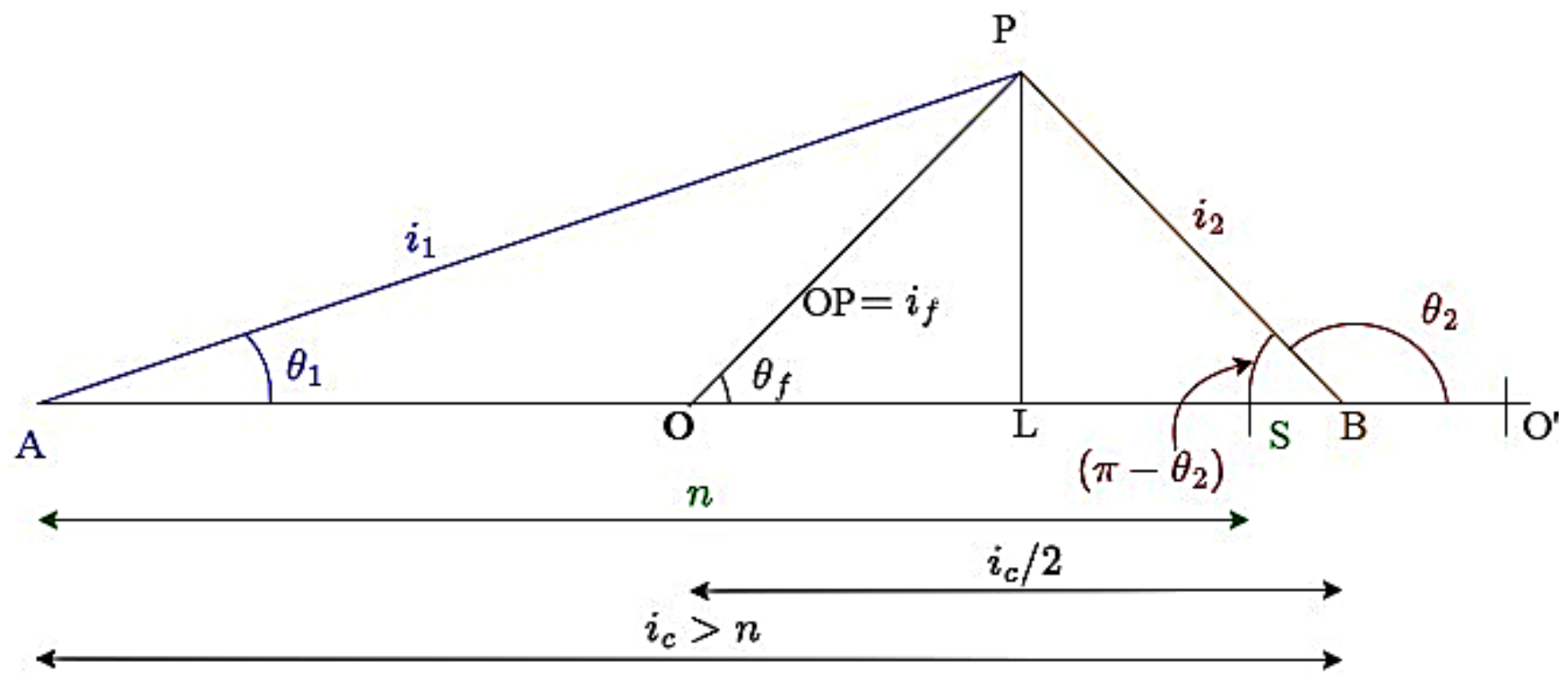

Figure S.6. Geometric description of a 2-chain junction. Description in the text.

$$\mathrm{AQ}^2 = \mathrm{QM}^2 + \mathrm{AM}^2 = \mathrm{OQ}^2 \sin^2\theta_f + (\mathrm{AO} + \mathrm{OM})^2 \tag{S.26}$$

$$n^2 = i_{fm}^2\left(1 - \cos^2\theta_f\right) + \left(\left(i_c/2\right) + i_{fm}\cos\theta_f\right)^2 \tag{S.27}$$

From the geometry, and the quadratic equation (S.27) solution, $i_{fm}$ and $i_2^0$, relate to $\cos\theta_f$ and $i_c$.

$$i_{fm} = \frac{1}{2}\left(-i_c \cos\theta_f + \sqrt{\left(i_c \cos\theta_f\right)^2 + 4\times\left[n^2 - \left(i_c/2\right)^2\right]}\right) \tag{S.28}$$

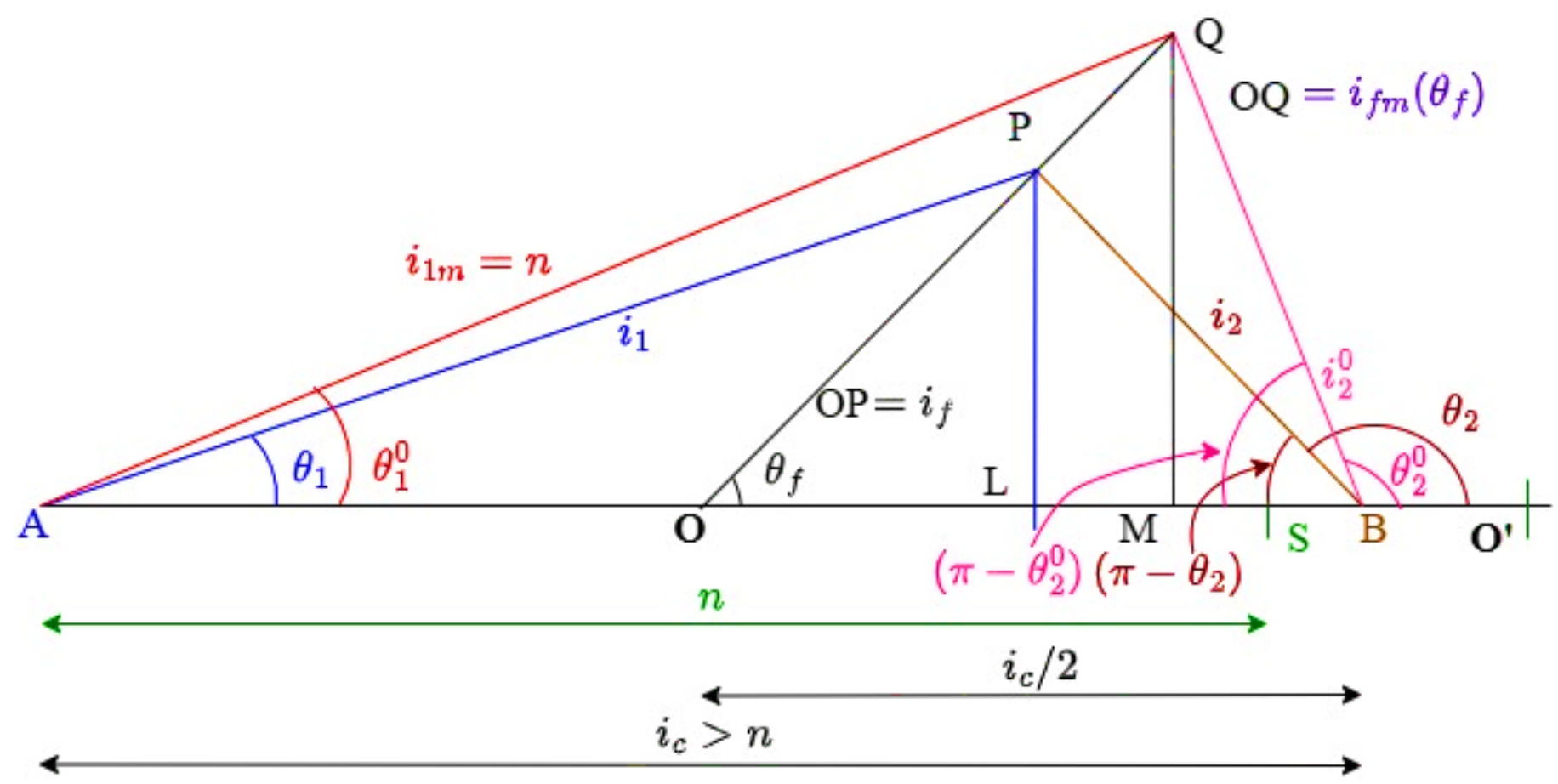

Figure S.7. At a given $\theta_f$, as AP=$i_1$ increases to AQ=$i_{1m} = n$ ($\theta_1$ correspondingly increases to $\theta_1^0$), then OP=$i_f$ increases to OQ=$i_{fm}$. Then $i_2$ increases to $i_2^0$ and $\theta_2$ decreases to $\theta_2^0$.

$$\left(i_2^0\right)^2 = \left(\left(i_c/2\right) - i_{fm}\cos\theta_f\right)^2 + i_{fm}^2\left(1-\cos^2\theta_f\right) \tag{S.29}$$

Given $i_c$, in Figure S.8, the semi-circle arc, U'P'PU, traces at a fixed $i_f$, the variation of $\theta_f$, in terms of $i_1$ and $i_2$. Different combinations of $i_1$ and $i_2$, yield corresponding values of $i_f$ and $\theta_f$; e.g., chains AP' (length $i'_1 = i_2$) and BP' (length $i'_2 = i_1$), meeting at P', are mirror images of the AP and BP chain combination (Figure S.6). Hence, the $\theta_f$ range from 0 to $\pi/2$, has a degeneracy of 2.

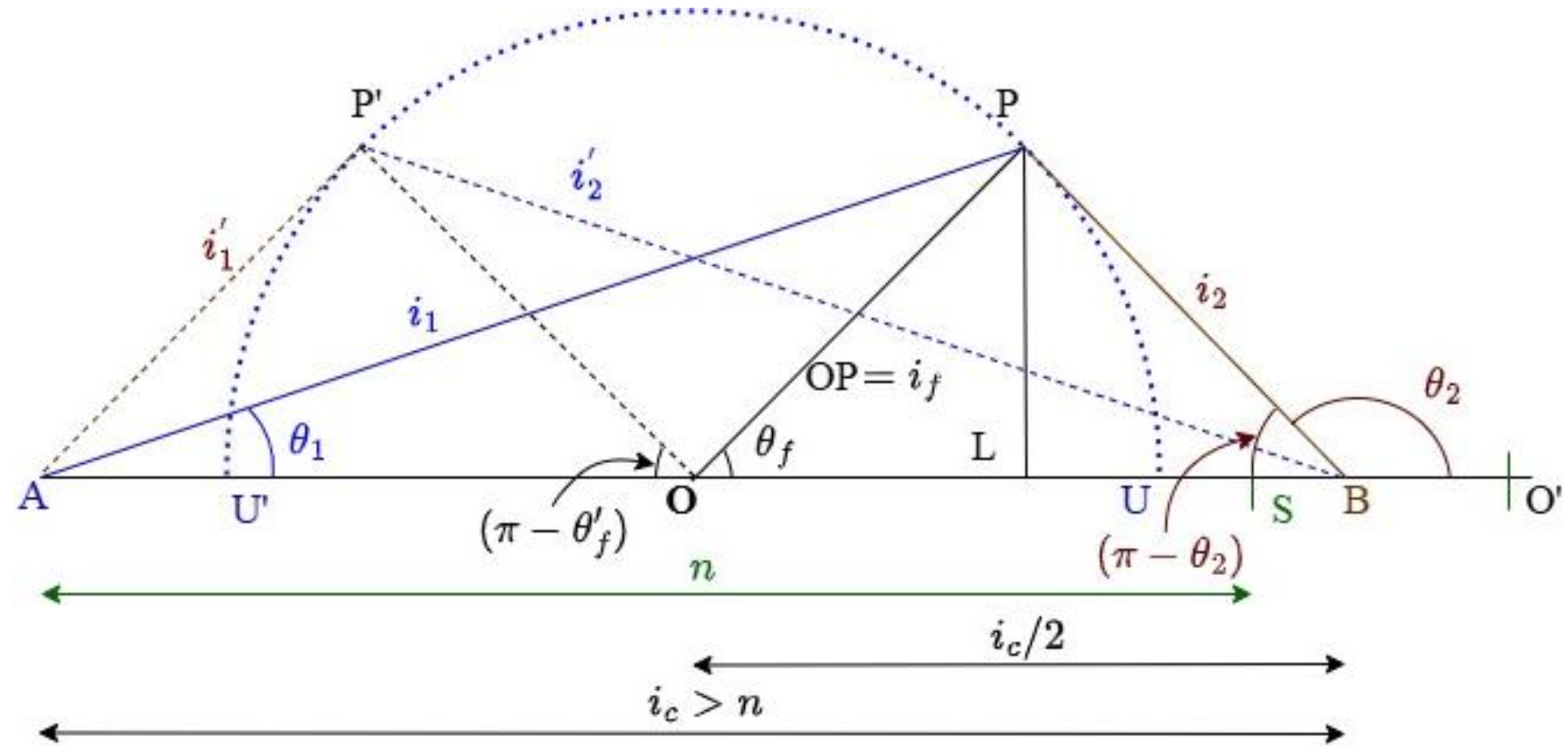

Figure S.8. Variation of $\theta_f$ in terms of $i_1$ and $i_2$, for a fixed $i_f$, and a special case of the mirror image of $i_1$ and $i_2$ chains, meeting at point P', on the semi-circle, U'P'PU, of radius $i_f$.

Figure S.9 describes the variation of both, $\theta_f$ (fixed $i_f$ ) and $i_f$ (fixed $\theta_f$ ) (Figure S.6 to Figure S.8), given $i_c$ . At $\theta_f = \pi/2$, $i_{fm}$ = OR and $i_2^0 = n$ =BR (not shown). The governing geometry of junction fluctuations, provides their distribution framework.

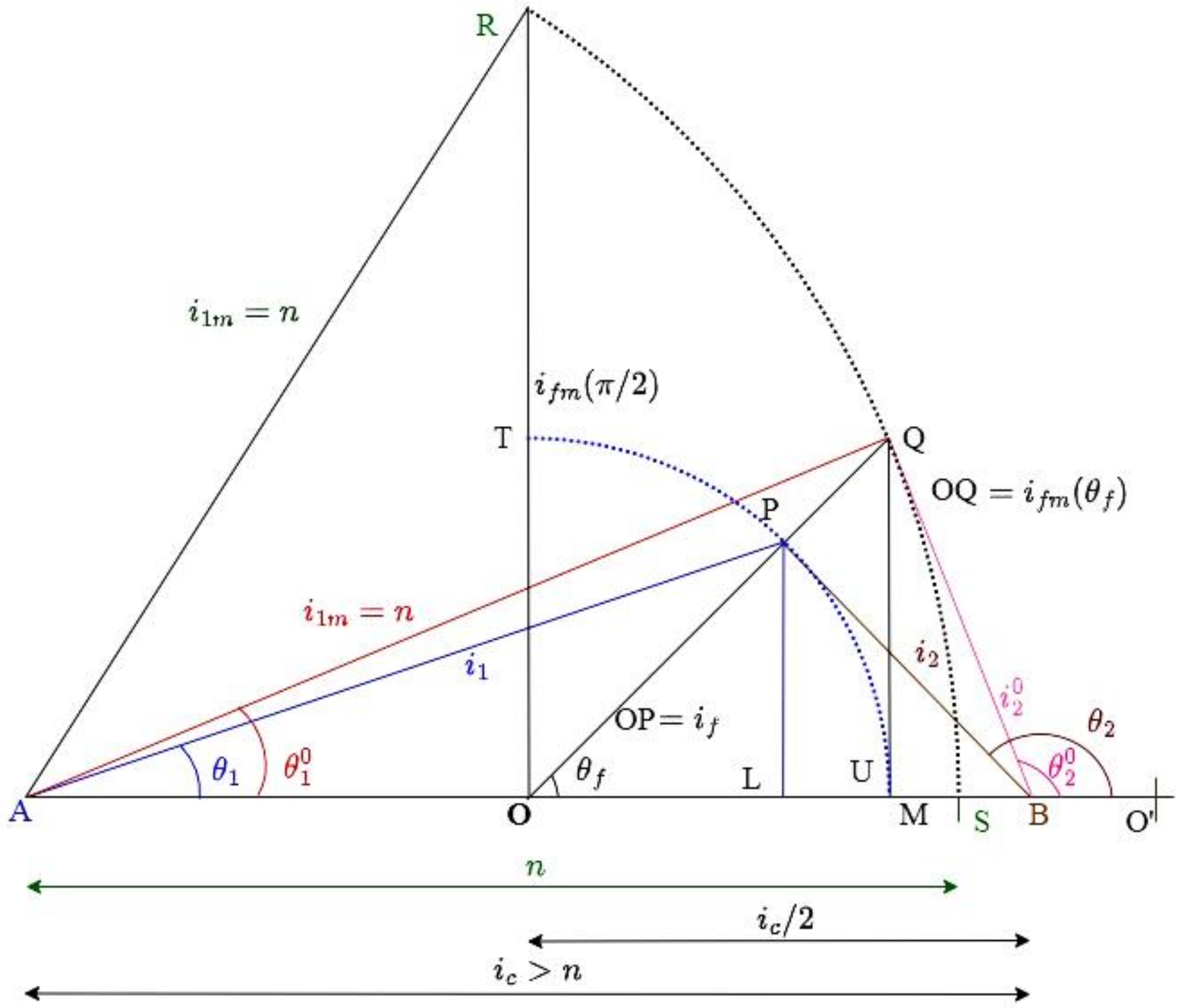

Figure S.9. The fluctuation geometry and ranges, summarized for a junction of two 3-D FJCs. Given $i_c$ , this figure describes the variation of both, $\theta_f$ up to $\theta_f = \pi/2$, and of $i_f$ up to $i_{fm}$ =OR.

### S.1.2.2 Fluctuation Distributions of Junctions of Two 3-D FJCs

The probability density, $\omega_f(i_f)$, of junction fluctuations (location of P), is the combined probability density, $\omega_n(i_1)$ , that the *n*-segment FJC from A (vector length $i_1$ ), and $\omega_n(i_2)$ , that the *n*-segment FJC from B (vector length $i_2$ ), end at P, given $\omega_{2n}(i_c)$, for a 2*n*-segments FJC, with distance AB= $i_c$ .

$$\omega_f(i_f) = \frac{\omega_n(i_1)\omega_n(i_2)}{\omega_{2n}(i_c)} \tag{S.30}$$

The three single-chain probability densities on the RHS, can be expressed in terms of the *m*=1 model (eqn. (S.12)) or the *m*=3 model (eqn. (S.13)) for equation (S.3). $i_f$ and $\theta_f$ , are coupled. The total elemental volume, accounting for the degeneracy of $\omega_f$ (within $d\omega_f$ , in the volume where $i_f$ varies within $di_f$ and $\theta_f$ varies within $d\theta_f$) is $dV_{\text{tot}} = 2\times i_f d\theta_f 2\pi i_f \sin\theta_f di_f$. There are two strips of radius $i_f \sin\theta_f$, width $i_f d\theta_f$ and thickness $di_f$, as shown in Figure S.10. $0 \le \theta_f \le \pi/2$.

$$\int_0^{i_{fm}(\theta_f)} \int_0^{\pi/2} P(i_f, \theta_f) d\theta_f di_f = 2\int_0^{i_{fm}(\theta_f)} \int_0^{\pi/2} \omega_f(i_f, \theta_f) i_f d\theta_f 2\pi i_f \sin\theta_f di_f = 1 \tag{S.31}$$

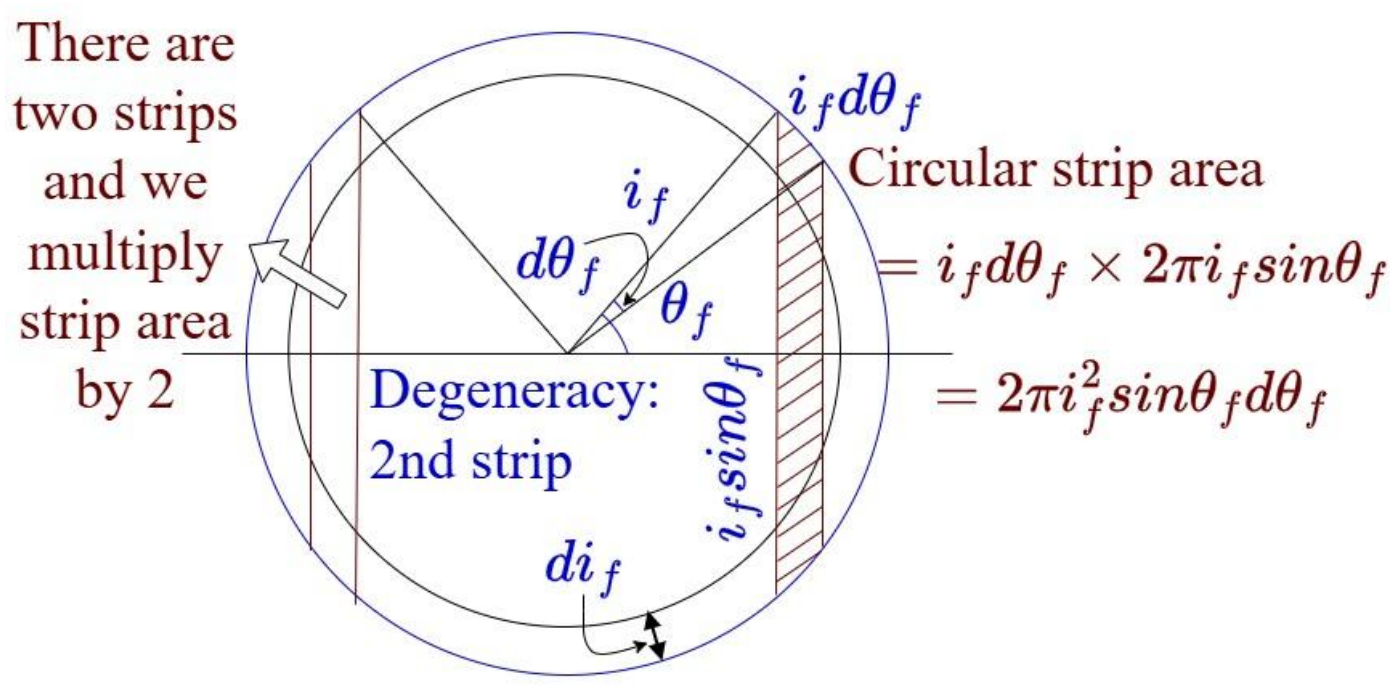

Figure S.10. Fluctuation $i_f$ degeneracy for a given $\theta_f$.

$i_f$ and $\theta_f$ depend on $i_1$, $i_2$, $\theta_1$ and $\theta_2$. The respective single chain $\omega_n(i_1)$, $\omega_n(i_2)$ and $\omega_{2n}(i_c)$, are functions of even powers of $i_1$, $i_2$ and $i_c$ (via eqns. (S.12) or (S.13)); we integrate over $i_f$ first.

$$\int_0^1 \int_0^{i_{fm}(\cos\theta_f)} 4\pi i_f^2 \omega_f di_f d(\cos\theta_f) = 1 \tag{S.32}$$

As the far ends of the two finite, ideal FJCs move apart ($i_c$ increases), the junction fluctuations are naturally restricted ($\theta_1^0$ decreases). This progressive suppression of fluctuations, quantitatively maps via two parameters to distributions of ideal FJCs of $i_c$-dependent $n_{f2}$ segments.

1. Comparison with $\Omega_0$ (OP=$i_f$=0)

The two chains from A and B, both terminate at O; i.e., $i_1 = i_2 = \left(i_c/2\right)$. For a two-chain junction, $n_{f2}$ is obtained from equations (S.33) and (S.34).

$$\omega_{nf}\big|_{(i_f=0)} = \frac{\omega_n\big|_{\left(i_1=i_c/2\right)} \omega_n\big|_{\left(i_2=i_c/2\right)}}{\omega_{2n}\big|_{(i_c)}} = \frac{\left(\omega_n\big|_{\left(i_c/2\right)}\right)^2}{\omega_{2n}\big|_{(i_c)}} \tag{S.33}$$

$$\Omega_{nf}\big|_{H_f=0} = (n_{f2} l)^3 \omega\big|_{(i_f=0)} = \frac{\Omega_{0G}}{1+\dfrac{0.75}{n_{f2}}\exp\left(\dfrac{0.5}{n_{f2}}\right)} \tag{S.34}$$

For various $n$ values, from Figure S.11a, normalized $\overline{n_f} = \dfrac{n_{f2}\left(\omega_{nf}\big|_{H_f=0}\right)}{\left(n/2\right)} = 1 - H_c^2$; $H_c^2 = \left(\dfrac{i_c}{2n}\right)^2$. For all $n$, $n_{f2}\big|_{i_c=0} = n/2 = n_{f2G}$ ($n_f$ of the junction fluctuation of two Gaussian chains) and linearly decreases with $H_c^2$ to $n_{f2}\big|_{i_c=2n}$ =0. Since equation (S.6) is valid for $n \geq 12$, the error in applying equation (S.34) progressively increases as $n_{f2}$ progressively decreases below 12, making the plot slightly nonlinear in the range, $0.5 < H < 1$.

2. Comparison with $\left\langle i_f^2 \right\rangle$

For an ideal FJC, $\left\langle i_f^2 \right\rangle = n_{f2}$.

$$\left\langle i_f^2 \right\rangle = \int_0^1 \int_0^{i_{fm}(\cos\theta_f)} i_f^2 \omega_f 4\pi i_f^2 di_f d\left(\cos\theta_f\right) \tag{S.35}$$

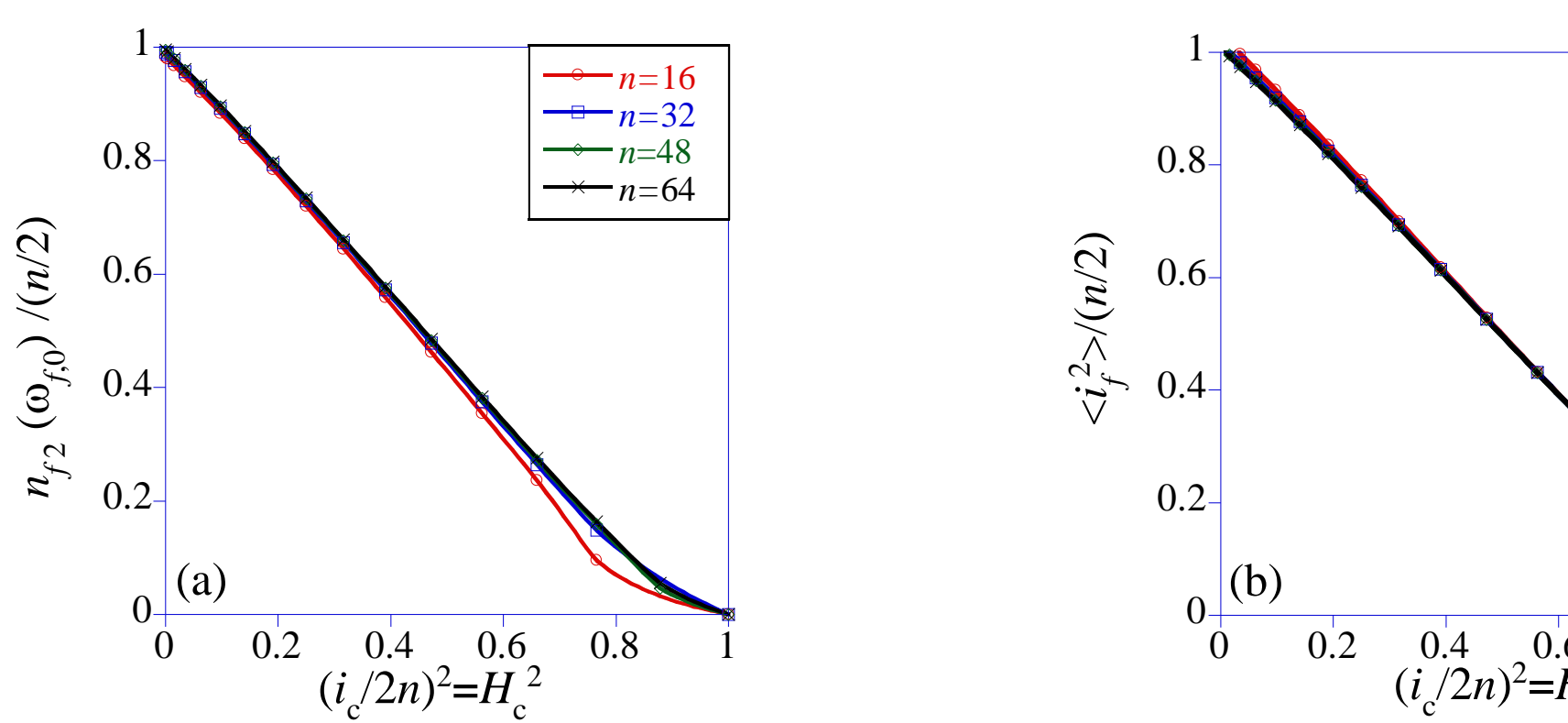

Figure S.11. Normalized junction fluctuations $\overline{n_f} = \dfrac{n_{f2}}{\left(n/2\right)} = 1 - H_c^2$; $H_c^2 = \left(\dfrac{i_c}{2n}\right)^2$. (a) Based on $\omega_{f,0}$ (b) Based on $n_{f2} = \left\langle i_f^2 \right\rangle$. Lines are to guide the eye.

Figure S.11b depicts that normalized $\overline{n_f} = \dfrac{n_{f2}\left(\left\langle i_f^2 \right\rangle\right)}{\left(n/2\right)} = 1 - H_c^2$. Both parameters of $\omega_f\left(i_f\right)$, $\omega_{f0}$ and $\left\langle i_f^2 \right\rangle$, correspond closely to those of single FJCs of $n_{f2}$ segments. $\overline{n_f} = n_{f2} \big/ \left(n/2\right) = 1 - H_c^2$.

We compare next, fluctuation distributions, $\omega_f$, with distributions, $\omega_{nf}$, of ideal FJCs of $n_{f2}$ segments. Both, $i_c$ and $\cos\theta_f$ (eqn. (S.28)), affect the fluctuation range. For $i_c$ =0 (Figure S.12a), at all $\cos\theta_f$, $n_{f2} = 0.5n$; fluctuation $h_{f\max} = 2n_{f2} = n$ (FJC $h_{\max}(n_{f2}) = n_{f2}$). $H_f = h_f / n_{f2}$; $H_{f\max} = 2$. We use mixed notation, $\omega(H)$. $\omega_{nf}\big|_{H=1} = 0$ (FJC); i.e., $\ln\omega_{nf}\big|_{H\to 1}$ diverges (not shown). For finite, normalizing, we choose $\ln\omega_{nf}\big|_{H\sim 0.99}$. From plots of $\ln\omega_{nf} \Big/ \left|\left(\ln\omega_{nf}\big|_{0.99}\right)\right|$ and $\ln\omega_f \Big/ \left|\left(\ln\omega_{nf}\big|_{0.99}\right)\right|$, $\omega_{f0} \sim \omega_{nf0}$. For $i_{\mathrm{c}} = n$ (Figure S.12b.), $H_{\mathrm{c}} = 0.5$; $\overline{n_f} = \dfrac{n_{f2}}{\left(n/2\right)} = 1 - H_{\mathrm{c}}^2 = 0.75$; for all $\cos\theta_f$, $\omega_{f0} \sim \omega_{nf0}$; $h_{f\max}\big|_{\cos\theta_f = 1} = i_{\mathrm{c}}/2 = (4/3)n_{f2} = 0.5n$; $h_{f\max}\big|_{\cos\theta_f = 0} = \left(4/\sqrt{3}\right)n_{f2} \sim 0.866n$. Thus, for both $i_{\mathrm{c}}$ values, the fluctuation ranges, $i_{fm} \geq n_{f2}$; i.e., the *distributions* are not *of* ideal FJCs.

Two-chain junction fluctuations map parametrically to FJCs. To tractably extend this analysis to actual networks, with junction functionality, $\phi > 2$, we first validate the comparison with two-chain junction fluctuations of 1-D FJCs and then, proceed via the four-chain junction fluctuations of 1-D FJCs.

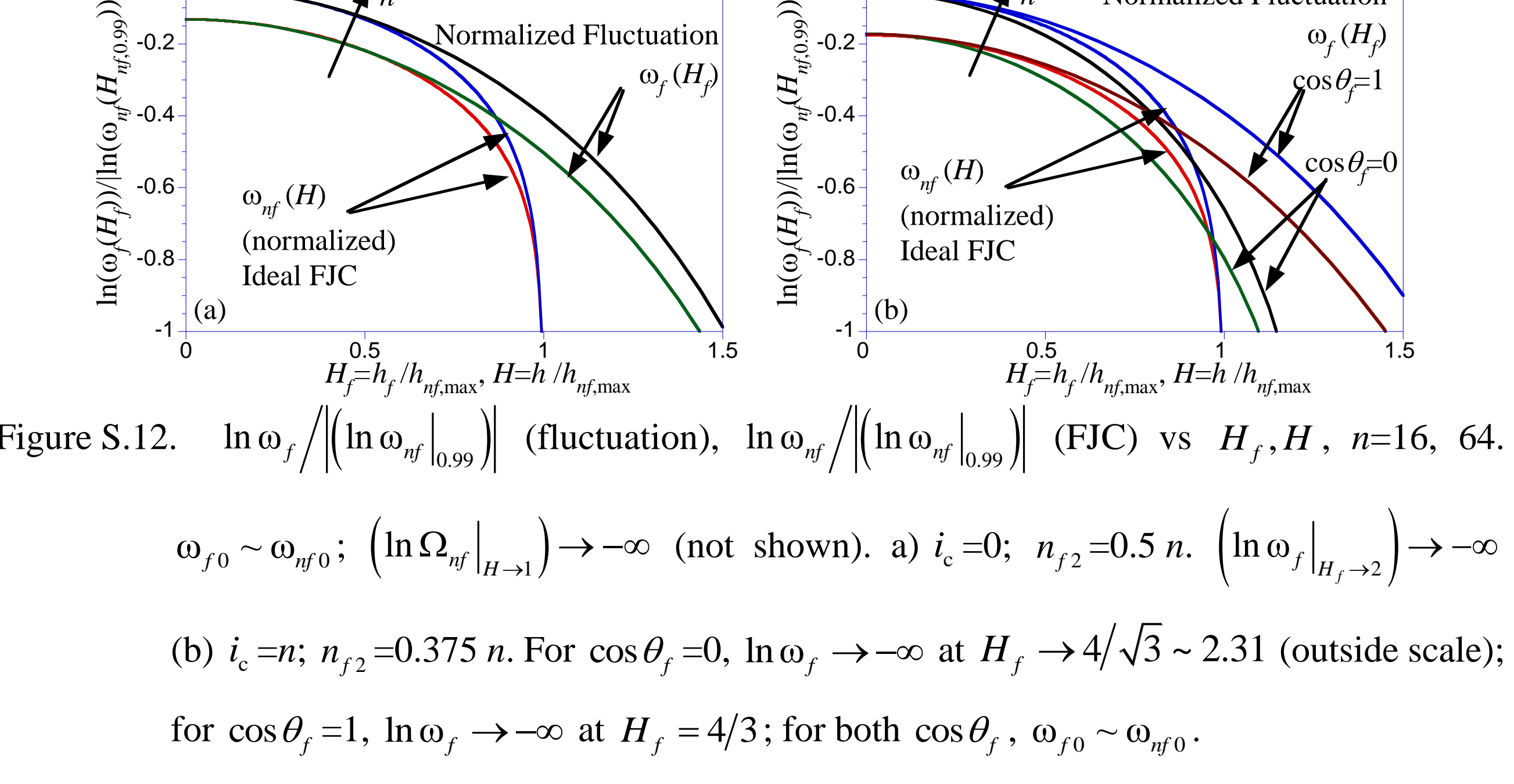

Figure S.12. $\ln\omega_f \Big/ \left|\left(\ln\omega_{nf}\big|_{0.99}\right)\right|$ (fluctuation), $\ln\omega_{nf} \Big/ \left|\left(\ln\omega_{nf}\big|_{0.99}\right)\right|$ (FJC) vs $H_f, H$, $n$=16, 64. $\omega_{f0} \sim \omega_{nf0}$; $\left(\ln\Omega_{nf}\big|_{H\to 1}\right) \to -\infty$ (not shown). a) $i_{\mathrm{c}}$=0; $n_{f2}$=0.5 $n$. $\left(\ln\omega_f\big|_{H_f\to 2}\right) \to -\infty$ (b) $i_{\mathrm{c}}$=$n$; $n_{f2}$=0.375 $n$. For $\cos\theta_f$=0, $\ln\omega_f \to -\infty$ at $H_f \to 4/\sqrt{3} \sim 2.31$ (outside scale); for $\cos\theta_f$=1, $\ln\omega_f \to -\infty$ at $H_f = 4/3$; for both $\cos\theta_f$, $\omega_{f0} \sim \omega_{nf0}$.

### S.1.2.3 Fluctuation of a Two-Chain Junction of 1-D FJCs

Analogous to the 3-D FJCs in Figure S.5, we consider two $n$-segment 1-D FJCs, one each from points A and B (Figure S.13), where the segments $(l = 1)$ can only be oriented left or right. The outer ends, A and B, of the two FJCs, are distance $i_{\mathrm{c}}$ apart. The junction is point P, the mid-point of the contour AB, the 1-D combined $2n$-segment FJC. Its mean position is at O, the mid-point of vector length $i_{\mathrm{c}}$.

For any $i_c$ varying from 0 to $2n$, $\omega_{n,\mathrm{1D}}(i_1)$ is the dimensionless vector probability that the FJC from A terminates at P; AP=$i_1$. $i_1$ can be positive (P is to the right of A) or negative (P is to the left of A) $\left(P_{n,\mathrm{1D}}=2\omega_{n,\mathrm{1D}}\right)$. Similarly, $\omega_{n,\mathrm{1D}}(i_2)$, describes the $i_2$ FJC from B. Positive $i_2$ is directed toward the right of B, $i_c = i_1 - i_2$. $\omega_{2n,\mathrm{1D}}(i_c)$ is the probability that the combined chain vector length, AB=$i_c$.

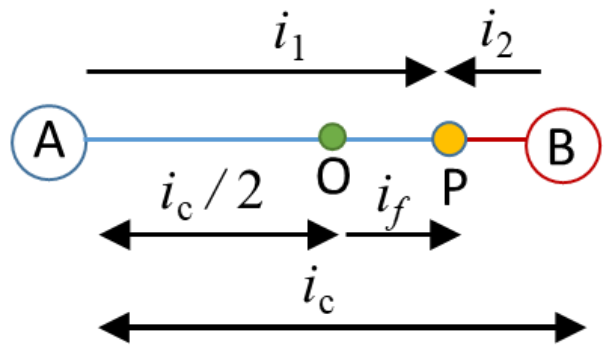

Figure S.13. Schematic of the geometry of junction fluctuations of two 1-D FJCs.

$$\omega_{n,\mathrm{1D}}(i_1)=\frac{n!}{\left(\frac{n-i_1}{2}\right)!\left(\frac{n+i_1}{2}\right)!}\frac{1}{2^n};\ \omega_{n,\mathrm{1D}}(i_2)=\frac{n!}{\left(\frac{n-i_2}{2}\right)!\left(\frac{n+i_2}{2}\right)!}\frac{1}{2^n};\ \omega_{2n,\mathrm{1D}}(i_c)=\frac{(2n)!}{\left(n-\frac{i_c}{2}\right)!\left(n+\frac{i_c}{2}\right)!}\frac{1}{2^{2n}}$$

Then, the vector probability that the two chains meet at P (distance $i_1$ from A and $i_2$ from B), given that for the outer ends of the chains, AB= $i_c$, $\omega_{\mathrm{P,1D}}=\dfrac{\omega_{n,\mathrm{1D}}(i_1)\,\omega_{n,\mathrm{1D}}(i_2)}{\omega_{2n,\mathrm{1D}}(i_c)}$. The maximum lengths of the $i_1$ FJC and the connected $i_2$ FJC can be $n$; i.e., the values of $i_1$ range from $i_c - n$ to $n$; $i_2 = i_1 - i_c$. Then, given $i_c$, $\omega_{f,\mathrm{1D}}=\omega_{\mathrm{P,1D}}$ is the distribution of the fluctuation of the junction P from the mean location, O. The fluctuating length, $i_f = i_1-(i_c/2)$; $i_f$ ranges from $(i_c/2)-n$ to $n-(i_c/2)$. For $i_c$ =0, $i_f$ ranges with $i_1$, from $-n$ to $+n$. As with the two-chain junction fluctuation distributions of 3-D FJCs, Figure S.14 plots $n_{f2}$ of $\omega_{f,\mathrm{1D}}$ as function of $i_c$, in terms of its parameters (i) $\omega_{0,\mathrm{1D}}$ and (ii) $\left\langle i_f^2\right\rangle$

1. Comparison with $\omega_{0,\mathrm{1D}}$

The probability of a single 1-D FJC is maximum at $i=0$ ($\omega_{0,\mathrm{1D}}$). If $n$ is an even number:

$$\omega_{0,\mathrm{1D}}=\frac{n!}{2^n\left(\left(\frac{n}{2}\right)!\right)^2} \tag{S.36}$$

$\omega_{\mathrm{G,1D}}=\omega_{\mathrm{0G,1D}}\exp\left(-\dfrac{nH^2}{2}\right)$, $\omega_{\mathrm{0G,1D}}=(2n\pi)^{-(1/2)}$*. $\Omega_{\mathrm{0G,1D}}=\left(n/(2\pi)\right)^{(1/2)}$. As with 3-D FJCs:

* Although $\omega_{0,\mathrm{1D}}$ is a dimensionless probability for finite $n$; in the limit of large $n$, it is expressed as the probability distribution, which has units of reciprocal length.

$$\omega_{0,1\mathrm{D}} = \frac{\omega_{0\mathrm{G},1\mathrm{D}}}{1+\frac{0.25}{n}\exp\left(\frac{0.0715}{n}\right)} \tag{S.37}$$

Equation (S.37) is analogous to equations (S.6) and (S.34). Given $i_\mathrm{c}$, $\omega_{f0} = \omega_f\big|_{i_f=0}$ ($i_1 = i_2 = (i_\mathrm{c}/2)$). Then, analogous to combining equations (S.33) and (S.34) (Figure S.11a) for 3-D FJCs, $n_{f2} = n$ in equation (S.37) with $\omega_{0,1\mathrm{D}} = \omega_{f0}$.

2. Comparison with $\left\langle i_f^2 \right\rangle$

For the second method to estimate $n_{f2}$, given $i_\mathrm{c}$, the mean squared length for a 1-D $n$-segments FJC, $\left\langle i^2 \right\rangle = \sum_{i=-n}^{n} i^2\omega_n(i) = \sum_{i=-n}^{n} i^2 \frac{n!}{\left(\frac{n-i}{2}\right)!\left(\frac{n+i}{2}\right)!}\frac{1}{2^n} = n = \left\langle i_f^2 \right\rangle\Big|_{i_\mathrm{c}} = \sum_{i_f=(i_\mathrm{c}/2)-n}^{n-(i_\mathrm{c}/2)} i_f^2\omega_f\left(i_f\right) \sim n_{f2}$; this is analogous to the analysis of equation (S.35) and Figure S.11b, with 3-D two-FJC junctions.

As in the case of the 3-D two-chain junction fluctuations, both methods indicate that the two-chain junction fluctuation of 1-D FJCs, is naturally suppressed, on stretching the chains (Figure S.14); $\overline{n_f} = n_{f2}\Big/\left(n/2\right) = 1 - H_\mathrm{c}^2$. The normalized FJC displacement, $H = i/n_{f2}$, and the normalized fluctuation, $H_f = i_f/n_{f2}$. Mixed notation, i.e., $\omega_{nf,1\mathrm{D}}(H)$, is used. Figure S.15 depicts $\ln\left(\omega_{f,1\mathrm{D}}\right)$ normalized with $\left|\ln\left(\omega_{nf,1\mathrm{D}}\big|_{H=1}\right)\right|$, when $i_\mathrm{c}$ =0, and its comparison with an ideal 1-D FJC of $n_{f2}$=$n$/2 segments. Unlike for 3-D FJCs, for discrete 1-D FJCs' distributions, $\omega_{nf,1\mathrm{D}}\big|_{H=1} \neq 0$. Normalized, $\Omega_{nf,1\mathrm{D}}(H) = n_{f2}\omega_{nf,1\mathrm{D}}\left(i_f\right)$.

As with two-chain junction fluctuations of 3-D FJCs, the two-chain junction fluctuation distribution of 1-D FJCs, is also not an ideal 1-D FJC; the fluctuation ranges, $i_{fm} \geq n_{f2}$; e.g., at $i_\mathrm{c}$=0, $i_f$ varies from $-n$ to $+n$. In contrast, for an ideal FJC of $n_{f2}$ =$n$/2 segments, its $h$ would range from $-n/2$ to $+n/2$. $\omega_f\left(i_f\right)$, however, shares its parameters, $\left\langle i_f^2 \right\rangle$ and $\omega_{0,1\mathrm{D}}$, with an FJC.

Thus, the normalized 1-D two-chain junction fluctuation distributions and their parametric mapping to a 1-D FJC, are similar to the normalized 3-D two-chain junction fluctuations and their parametric mapping to a 3-D FJC. Next, we infer the four-chain junction fluctuation distributions for 3-D FJCs via four-chain junction fluctuations of 1-D FJCs, and hence, for junctions of any $\phi > 2$, of 3-D FJCs.

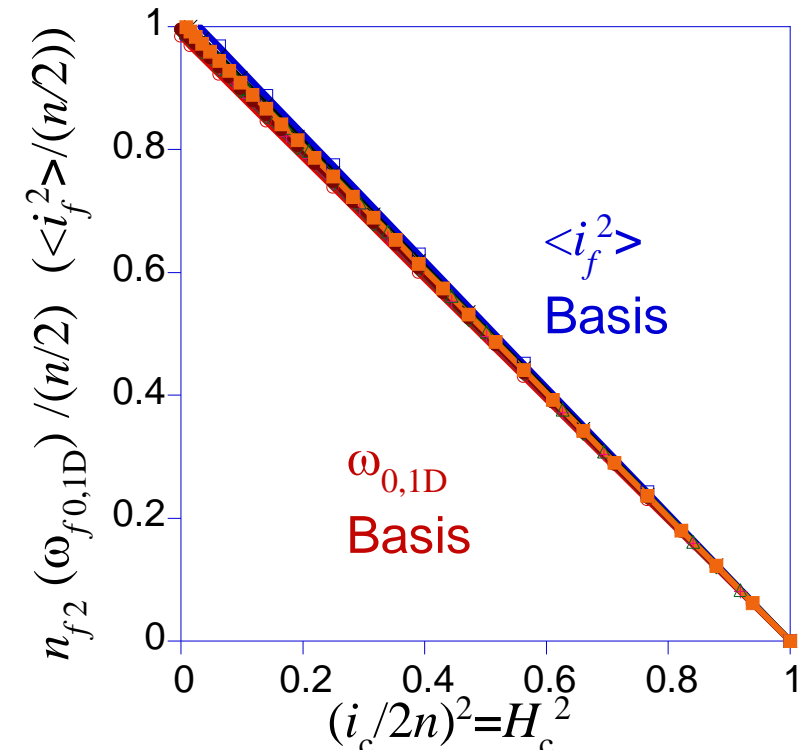

Figure S.14. Normalized $\overline{n_f}=\dfrac{n_{f2}\left(\omega_{f0,1D}\right)}{\left(n/2\right)}$ and $\overline{n_f}=\dfrac{n_{f2}\left(\left\langle i_f^2\right\rangle\right)}{\left(n/2\right)}$; $\overline{n_f}=1-H_c^2$; $H_c^2=\left(\dfrac{i_c}{2n}\right)^2$, for two-chain junction fluctuations of 1-D chains. Lines are to guide the eye.

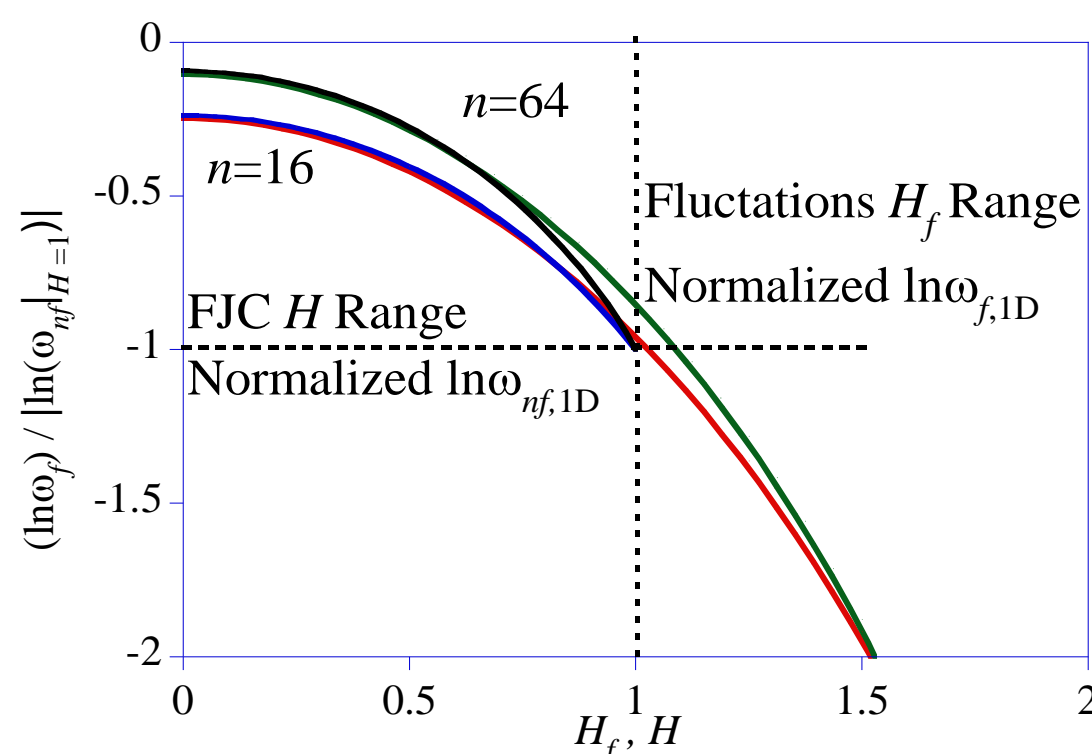

Figure S.15. $\dfrac{\ln\omega_{f,1D}}{\left|\ln\left(\omega_{nf,1D}\big|_{H=1}\right)\right|}$ (fluctuation) and $\dfrac{\ln\omega_{nf,1D}}{\left|\ln\left(\omega_{nf,1D}\big|_{H=1}\right)\right|}$ (FJC) vs $H_f, H$; $i_c$=0, $n_{f2}$=0.5$n$; $n$=16, 64 (+ve range shown). $\omega_{f0,1D}\sim\omega_{nf0,1D}$. Normalized FJC curves meet at $H=1$; Normalized $\ln\omega_{f,1D}$ curve range $H_f=2$ (beyond scale) (i.e., $i_c$=2$n$, $H_c$=1).

### S.1.2.4 Fluctuation Distribution of the Four-Chain Junction of 1-D FJCs

We consider four 1-D FJCs, meeting at a junction. Analogous to two-chain junctions of 1-D FJCs, (Figure S.13), the four chains are distributed as two chains from A and two chains from B (distance AB=$i_c$) $(l=1)$, meeting at junction P. The segment directions sequences (contours) of the two chains from A or B are independent, for the same vector lengths (AP and BP). O, the midpoint of AB, is the mean position of P. The junction fluctuation distribution is the distribution of OP, for $i_c$ varying from 0 to 2$n$. The probability that AP=$i_1$, is $\omega_n\left(i_1\right)$. Similarly, $\omega_n\left(i_2\right)$ is the probability that BP=$i_2$. $i_1$ can be positive (P is to the right of A) or negative (P is to the left of A). $i_2$ is positive to the right of B; $i_2=i_1-i_c$. The probability that the combined chain is of vector length $i_c$, is $\omega_{2n}\left(i_c\right)$.

$$\omega_n(i_1)=\frac{n!}{\left(\frac{n-i_1}{2}\right)!\left(\frac{n+i_1}{2}\right)!}\frac{1}{2^n};\ \omega_n(i_2)=\frac{n!}{\left(\frac{n-i_2}{2}\right)!\left(\frac{n+i_2}{2}\right)!}\frac{1}{2^n};\ \omega_{2n}(i_c)=\frac{(2n)!}{\left(n-\frac{i_c}{2}\right)!\left(n+\frac{i_c}{2}\right)!}\frac{1}{2^{2n}}$$

The probability that the two FJCs from A have vector length AP=$i_1$, the two FJCs from B have vector length BP=$i_2$, given that the two combined FJCs, each of 2$n$ segments, have vector length AB= $i_c$, is $\omega_{P,1D}=\frac{(\omega_{n,1D}(i_1))^2(\omega_{n,1D}(i_2))^2}{(\omega_{2n,1D}(i_c))^2}$. The ranges of $i_1$ and $i_2$ are the same as in Section S.1.2.2, i.e., $i_c-n$ to $n$ for $i_1$ with $i_2=i_1-i_c$.

The fluctuating length, OP, is $i_f=i_1-(i_c/2)$, and its range is, $(i_c/2)-n$ to $n-(i_c/2)$. As in Section S.1.2.2, $\omega_P=\omega_f$. $n_{f4}$ of four-chain junction fluctuations of 1-D FJCs, is a function of $i_c$, (Figure S.16) via two parameters, (i) $\omega_{0,1D}$ and (ii) $\langle i_f^2\rangle$. These junction fluctuations are naturally suppressed as the chains extend; $n_{f4}$=$n$/4 segments when $i_c$=0, and $n_{f4}$=0, when $i_c$=2$n$.

Figure S.17 depicts the normalized fluctuation distribution plots of $i_f$ at $i_c$=0, analogous to those in Section S.1.2.2. $\omega_f$ parametrically maps to $\omega_{nf}$, the distribution of an ideal FJC of $n$/4 segments. The vector length range for such an ideal FJC is from $-n/4$ to +$n$/4, while the $i_f$ range is from $-n$ to +$n$. Hence, for the 4-chain junction also, the equivalent fluctuating chain of $n_{f4}$ segments, is not an ideal FJC, while it also shares distribution parameters, $\omega_{0,1D}$ and $\langle i_f^2\rangle$, with an FJC. The fluctuation distributions of two-chain junctions of ideal 1-D FJCs follow those of two-chain junction fluctuations of ideal 3 - D FJCs (Section S.1.2.3). Then, by analogy, the fluctuation distribution of four-chain junctions of 3 - D FJCs would follow that of four-chain junctions of 1-D FJCs (network junction functionality, $\phi$=4), $\overline{n_f}=n_{f4}/(n/4)=1-H_c^2$. Then, for any junction functionality, $\phi$, $\overline{n_f}=n_{f\phi}/(n/\phi)=1-H_c^2$.

As per an earlier development [26], the fluctuation of the junction of $\phi$ Gaussian chains, each of $n$ segments, is the distribution of a Gaussian chain of $n_{f\phi}$ segments, where throughout the deformation, $n_{f\phi}$=$n$/$\phi$. For ideal junctions of ideal FJCs, the natural decrease of $n_{f\phi}$ during network deformation, manifests in the resulting network uniaxial stress-deformation $(\sigma-\lambda)$ relationship. We describe next, the resulting $\sigma-\lambda$ relationship for a network of ideal 3-D FJCs, with ideal junction fluctuations.

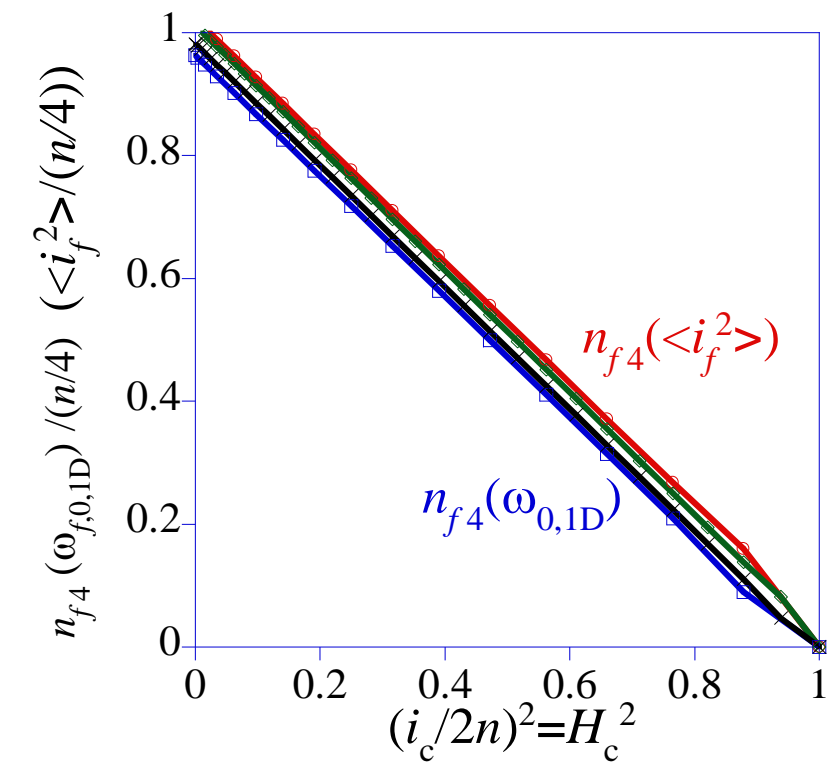

Figure S.16. Normalized $\overline{n_f} = \dfrac{n_{f4}\left(\omega_{f0,1D}\right)}{\left(n/4\right)}$ and $\overline{n_f} = \dfrac{n_{f4}\left(\left\langle i_f^2 \right\rangle\right)}{\left(n/4\right)}$; $\overline{n_f} = 1 - H_c^2$; $H_c^2 = \left(\dfrac{i_c}{2n}\right)^2$, for 4-Chain Junction Fluctuations for 1-D chains. Lines are to guide the eye.

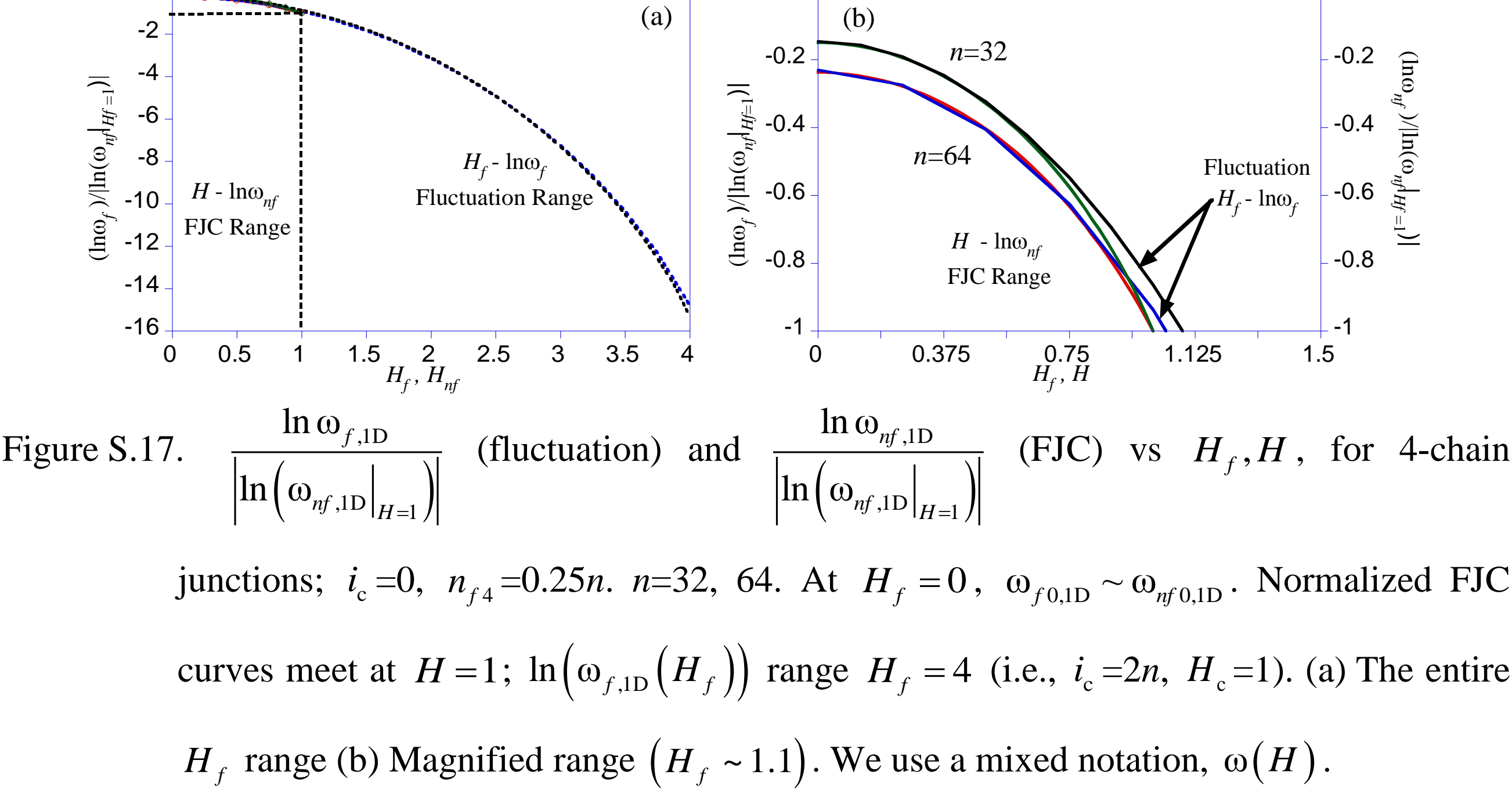

Figure S.17. $\dfrac{\ln\omega_{f,1D}}{\left|\ln\left(\omega_{nf,1D}\big|_{H=1}\right)\right|}$ (fluctuation) and $\dfrac{\ln\omega_{nf,1D}}{\left|\ln\left(\omega_{nf,1D}\big|_{H=1}\right)\right|}$ (FJC) vs $H_f, H$, for 4-chain junctions; $i_c$=0, $n_{f4}$=0.25$n$. $n$=32, 64. At $H_f = 0$, $\omega_{f0,1D} \sim \omega_{nf0,1D}$. Normalized FJC curves meet at $H = 1$; $\ln\left(\omega_{f,1D}\left(H_f\right)\right)$ range $H_f = 4$ (i.e., $i_c$=2$n$, $H_c$=1). (a) The entire $H_f$ range (b) Magnified range $\left(H_f \sim 1.1\right)$. We use a mixed notation, $\omega\left(H\right)$.

## S.2. Ideal Phantom Network

Section S.2.1 describes ideal FJC network deformation, with non-fluctuating junctions. Section S.2.2, then incorporates ideal junction fluctuations, to describe ideal phantom networks.

### S.2.1 Affine network of ideal FJCs – no junction fluctuations

Ideal FJC network elasticity relationships contrast with ILF-based chain developments [4,6], which have been extended further to networks (e.g., [4,32–41]). We describe first the ideal affine network deformation, which precludes junction fluctuations. $\lambda$, the bulk uniaxial deformation ratio of the deformed length to the undeformed length along the deformation axis, and the consequent radial deformations, are reflected identically in each of the network chains. The network relationships arise from the single-chain deformation work from Section S.1.1; i.e., equations (S.12) ($m$=1 model) and (S.13) ($m$=3 model), which are combined with the mathematical description of network geometry.

#### S.2.1.1 Chain Deformation as Function of Chain Orientation

We follow here, the previously described [35] axially symmetric deforming network geometry, in terms of the undeformed chain orientation (Figure S.18). In an isotropic undeformed network, the undeformed chain of length $h_0$, from the origin O, terminates at $P_0$, with cylindrical coordinates $(r_0, z_0)$. The undeformed chain vector $OP_0$, makes angle $\Theta_0$ to the deformation direction. $r_0 = h_0 \sin\Theta_0$ and $z_0 = h_0 \cos\Theta_0$. The angle $\alpha$ is not relevant, because uniaxial stress makes the deformed FJC network axially symmetric.

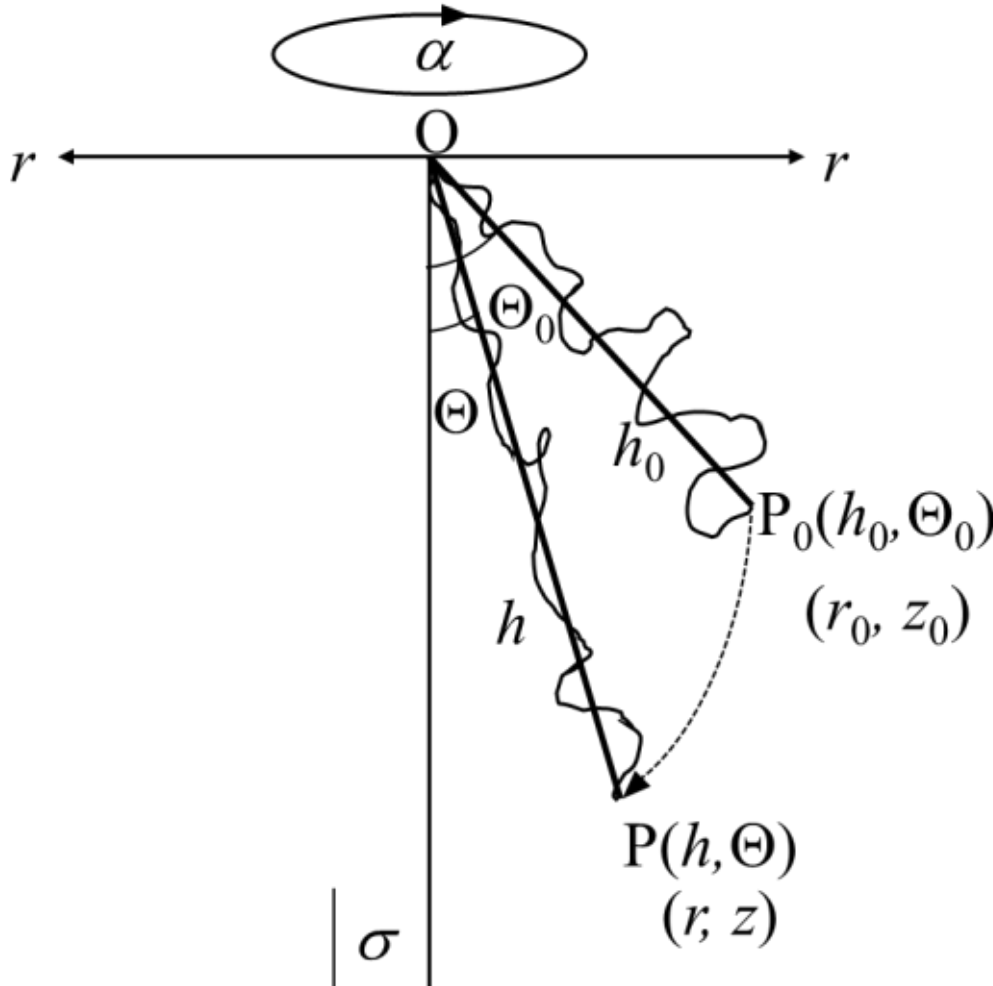

Figure S.18. Uniaxially symmetric geometry during uniaxial deformation [35] as function of undeformed chain length and orientation.

We suggest that, despite simulations that describe instantaneous network chain vector length polydispersity [42–49], the significantly longer deformation time-scale (compared to the chain fluctuation time scale) requires that the undeformed chains deform from the unperturbed (mean or most likely) vector lengths. Treloar has also argued [34] that all the undeformed chains in a crosslinked network, being of equal vector length, describes a real network, better than considering the vector length distribution of crosslinked chains to be that of isolated chains. Then, for all undeformed chains $H_0^2 = 1/n$ (eqn. (S.17)) and $H_0^{2t} = (1/n)^t$ (for $t > 1$). Network deformation results in chain vector re-orientation and deformation. The deformed chain terminates at $P(r, z)$. The vector, OP, is of length $h$, and makes an angle $\Theta$ with the deformation axis; $r = h\sin\Theta$ and $z = h\cos\Theta$. Incompressible $\left(r^2 z = r_0^2 z_0\right)$, affine uniaxial deformation ($\lambda$=$L$/$L_0$) yields, $z = \lambda z_0$ and $r_0^2 = \lambda r^2$.

At any $\Theta_0$, $\lambda\big|_{H\to 1} = \lambda_{\mathrm{m}}$, the extremum network $\lambda$ (maximum for tensile, minimum for compressive).

$\frac{H^2\big|_{\lambda_{\mathrm{m}}}}{H_0^2} = n$. If $\Lambda_n = \frac{h^2}{h_0^2} = nH^2$, $\Lambda_{n,\mathrm{m}} = \Lambda_n\big|_{\lambda_{\mathrm{m}}} = \left(\lambda_{\mathrm{m}}^2 \cos^2\Theta_0 + \frac{1}{\lambda_{\mathrm{m}}}\sin^2\Theta_0\right) = n$, and $\Lambda = \frac{\Lambda_n}{\Lambda_{n,\mathrm{m}}}$, then the deformed state, $H^2 = \Lambda$, and $\left(H^{2t} - H_0^{2t}\right) = \Lambda^t - \left(1/n^t\right)$, for any $t$.

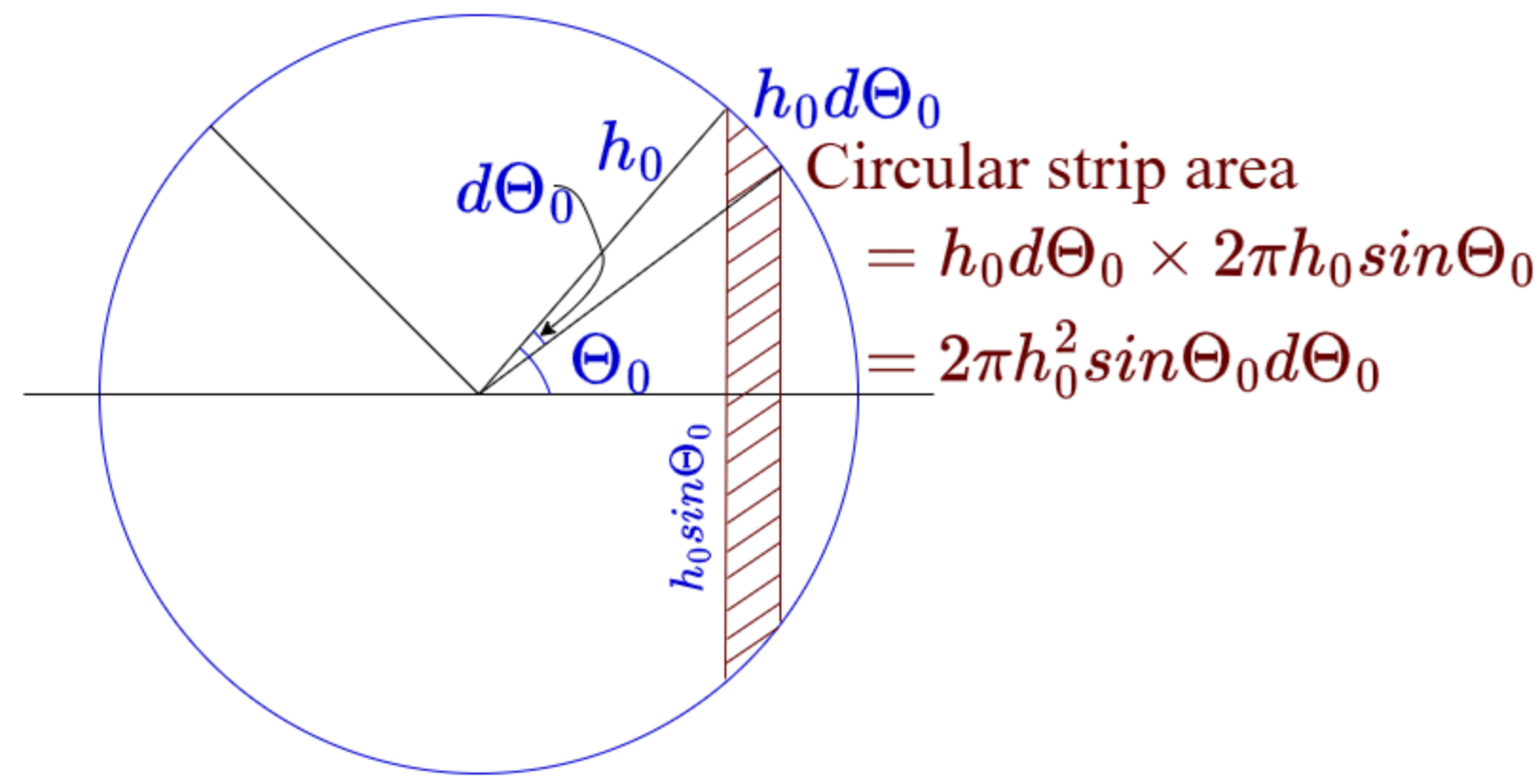

Figure S.19. Distribution of unperturbed ($h$=$h_0$) chain vector angle ($\Theta_0$) with the reference direction.

All undeformed chains are isotropically directed, ending on the surface of the sphere (surface area $= 4\pi h_0^2$). The number of chains within $d\Theta_0$ at $\Theta_0$, is proportional to $2\pi h_0^2\,\mathrm{Sin}\left(\Theta_0\right)d\Theta_0$, the area of the circular strip (Figure S.19). Dividing by $= 4\pi h_0^2$, $P\left(\Theta_0\right)d\Theta_0 = \frac{dN}{N} = \frac{\sin\left(\Theta_0\right)d\Theta_0}{2}$, the basis for integration over $\Theta_0$ [31,32,35,40,48–59]. We first consider the effect of $P\left(\Theta_0\right)d\Theta_0$.

Incompressible $\lambda_{\mathrm{m}}$ varies with $\Theta_0$, depending on whether the deformation is tensile or compressive, to be consistent with $\Lambda_{n,\mathrm{m}} = n$. Tensile $\lambda_{\mathrm{m}}\big|_{\cos\Theta_0 = 1} = n^{0.5}$, increases to $\lambda_{\mathrm{m}}\big|_{\cos\Theta_0 = 0} \to \infty$. Similarly, uniaxial compression $\lambda_{\mathrm{m}}\big|_{\cos\Theta_0 = 0} \to n^{-1}$, decreases to $\lambda_{\mathrm{m}}\big|_{\cos\Theta_0 = 1} \to 0$; i.e., there is a $\lambda_{\mathrm{m}}$ distribution within the network corresponding to $P\left(\Theta_0\right)d\Theta_0$; e.g., for tensile $\lambda^2 > n$, $\Theta_0$=0 chains can no longer deform.

For tetragonal solids, the cube body diagonal for a fixed volume, is the shortest, corresponding to the shortest vector length during deformation. All chains exhibit the same affine axial $\lambda$. During elongation, $\cos^2\Theta$ increases. All undeformed chains with $\cos^2\Theta_0 < \left(1/3\right)$ (flattened tetragon chains), contract until they reach $\cos^2\Theta = \left(1/3\right)$ (cube semi-diagonal); they extend when $\cos^2\Theta_0 \geq \left(1/3\right)$.

During uniaxial compression, $\cos^2\Theta$ decreases. Undeformed chains with $\cos^2\Theta_0 > (1/3)$ (extended tetragon chains), contract until they reach $\cos^2\Theta = (1/3)$; they extend when $\cos^2\Theta_0 \le (1/3)$.

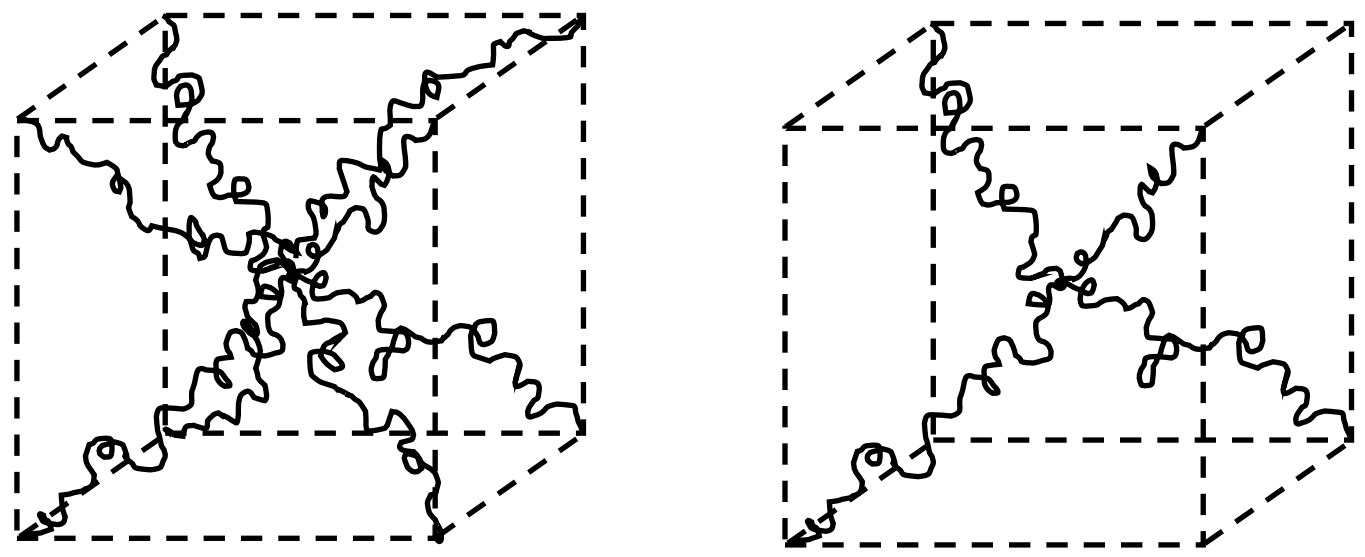

Figure S.20. BCC-type networks: Arruda Boyce 8-chain [41] and 4-chain network model [38,40].

During uniaxial extension or compression, the average and representative chains need to *extend*. Then the combined $\cos^2\Theta_0 = (1/3)$ and affine assumptions, preclude the inconsistencies of isotropic $P(\Theta_0)d\Theta_0$. All chains consistently exhibit uniform $\lambda_{\mathrm{m}}$, via affine and incompressible deformation; chain entropy would uniformly decrease since all chain vectors would extend monotonically during uniaxial tension or compression. Also, whether we integrate over $\cos\Theta_0$, or consider uniform $\cos^2\Theta_0 = \langle\cos^2\Theta_0\rangle = 1/3$, $h^2 = \frac{h_0^2}{3}\times\left(\lambda^2 + \frac{2}{\lambda}\right)$. $H^2 = \frac{H_0^2}{3}\times\left(\lambda^2 + \frac{2}{\lambda}\right)$; $\left(\lambda_{\mathrm{m}}^2 + \frac{2}{\lambda_{\mathrm{m}}}\right) = 3n$ [38,40,41].

This uniform $\cos\Theta_0$ is thus an equivalent geometry, with all undeformed chains in their "average", representative, orientation to the deformation axis. The integrated deformation over $\Theta_0$, would be consistent with the macroscopic incompressible deformation. Then, the junctions of the undeformed network, are at the center of a cube. Two versions of this BCC-type network, have been reported – the Arruda-Boyce 8-chain model [41] and its subset, the tetrahedral 4-chain model [38,40]. The chains originate from the center and reach the 8 corners or 4 of the 8 corners, respectively, of the cube (Figure S.20). The 4-chain model indicates that an "inclined" tetrahedron can be inscribed within a cube.

### S.2.1.2 Network Stress-Deformation Relationships

From above, the ideal FJC distribution provides $W_{def}(\Lambda) = W - W|_{H_0}$, the single-chain deformation work, representing the network relationships. For the $m$=1 and the $m$=3 models, respectively,

$$W_{def} = \left(\frac{n + 0.5n^{0.75}}{2}\right)\left(\Lambda - \frac{1}{n}\right) + \left(\frac{3 + 0.5n^{0.75} - 2n}{2}\right)\ln\left(\frac{1-\Lambda}{1-(1/n)}\right) \tag{S.38}$$

$$W_{def} = \frac{1}{2}\left(\frac{2}{n} + 1 + n\right)\left(\Lambda - \frac{1}{n}\right) + (2-n)\ln\left(\frac{1-\Lambda}{1-(1/n)}\right) - \frac{n}{20}\left(\Lambda^2 - \frac{1}{n^2}\right) - \frac{n}{15}\left(\Lambda^3 - \frac{1}{n^3}\right) \tag{S.39}$$

For a network of $N$ $n$-segment FJCs per unit volume, uniaxial strain energy density, $\mathbf{w} = N\mathrm{k}TW_{def}$, and stress, $\sigma = \dfrac{d\mathbf{w}}{d\lambda}$. If the network is defect-free, with no dangling chains, $N\mathrm{k}T = \dfrac{\rho \mathrm{R} T}{M_X}$; $\rho$ is the density, R is the universal gas constant, $T$ is the absolute temperature and $M_X = nM_s$ is the molecular weight of the chain between crosslinks; $M_s$ is the segment molecular weight.

Defining dimensionless $\mathbf{W} = \dfrac{\mathbf{w}}{(\rho \mathrm{R} T)/M_S}$ and $\boldsymbol{\Sigma} = \dfrac{d\mathbf{W}}{d\Lambda}$, for affine networks, $\mathbf{W}_{Af} = \dfrac{W_{def}}{n}$. Then

$$\frac{d\Lambda}{d\lambda} = \left(\lambda - \frac{1}{\lambda^2}\right)\frac{2}{3n} \tag{S.40}$$

$$\sigma_{Af} = \left(\lambda - \frac{1}{\lambda^2}\right)\frac{2}{3n}\frac{d\mathbf{w}_{Af}}{d\Lambda} = \left(\lambda - \frac{1}{\lambda^2}\right)\sigma_{red,Af} \tag{S.41}$$

For the $m$=1 model,

$$\begin{aligned}\sigma_{Af} &= \frac{\rho \mathrm{R} T}{nM_S}\frac{2}{3n}\left(\lambda - \frac{1}{\lambda^2}\right)\left(\frac{n + 0.5n^{0.75}}{2} + \frac{\left(3 + 0.5n^{0.75} - 2n\right)}{2(1-\Lambda)}\right) \\ &= \frac{\rho \mathrm{R} T}{nM_S}\frac{2}{3n}\left(\lambda - \frac{1}{\lambda^2}\right)\boldsymbol{\Sigma}_{Af}\end{aligned} \tag{S.42}$$

In case of the $m$=3 model,

$$\boldsymbol{\Sigma}_{Af} = \left(\frac{1}{2}\left(\frac{2}{n} + 1 + n\right) - \frac{n\Lambda}{10} - \frac{n\Lambda^2}{5} + \frac{(n-2)}{1-\Lambda}\right) \tag{S.43}$$

The significance of $\boldsymbol{\Sigma}$ and $\sigma_{red}$ will be discussed in Section S.2.2.2. Next, we describe the consequences of junction fluctuations on polymer network elasticity.

### S.2.2 Ideal Phantom Network

Each network chain is connected to the macroscopic surface by the path of chains, connecting the coordinated fluctuating junctions. Since the network junctions' mean positions move affinely during deformation, $H^2 = H_c^2$. Since $H^2 = \Lambda$ (Section S.1.2.2), $\overline{n_f} = 1 - \Lambda$. The consequent deformation-dependent junction effectiveness in network relationships, defines phantom networks.

#### S.2.2.1 Network Junction Fluctuations – Effective Ideal Chain to Phantom Network

To date, the analyses of how network junction fluctuations relax the network stresses from the affine deformation levels, have considered networks of Gaussian chains [17,20,21,23,25–28,30], and have resulted in an additional pre-factor of ½; i.e., the stress for uniaxial deformation $\lambda$, reduces from $\frac{\sigma}{N\mathrm{k}T}=\left(\lambda-\frac{1}{\lambda^2}\right)$ to $\frac{\sigma}{N\mathrm{k}T}=\frac{1}{2}\left(\lambda-\frac{1}{\lambda^2}\right)$. In addition, this pre-factor has been considered to remain deformation-invariant (consistent with invariant $n_f$), even in models for networks of finitely extensible chains – for models based on the ILF [18,19,24,29,31,33], FJC [31], or based on finite extensibility factors [22].

To describe the geometry of bulk network junction fluctuations, we trace the path of the fluctuating junctions of the connected chains, from the macroscopic non-fluctuating surface, up to each observed network chain in the bulk. Several groups of $\phi-1$ inward-directed, $n$-segment chains (called zero-seniority chains [26]), emanate from the outside boundary of the bulk (i.e., the macroscopic surface) (see Figure S.21 (a)). Each group meets at a junction. There are several such junctions at the first "layer" away from the boundary. Each chain-group junction, parametrically maps to a single sub-chain of $n_{f(\phi-1)}=\left(n_{f(\phi-1)\mathrm{G}}\overline{n_f}\right)=n_1\overline{n_f}=\frac{n\overline{n_f}}{\phi-1}=ny$ segments; the previously defined [26] $n_1=n_{f(\phi-1)\mathrm{G}}$ term, combines with the deformation dependent junction fluctuation effect, $\overline{n_f}$ (section S.1.2). There is a vacancy for one network chain at each junction, and a seniority-1 network chain emanates from each first-layer junction. The sub-chain of $n_1\overline{n_f}$ segments from the surface, "adds" to the connected seniority-1 network chain. The resultant combined seniority-1 chain contains $n_1\overline{n_f}+n=n\left(y+1\right)$ segments. Several groups of $\phi-1$ combined seniority-1 chains, meet at a junction each (Figure S.21 (b)), to form sub-chains of $n_2\overline{n_f}=\frac{\left(n_1\overline{n_f}+n\right)\overline{n_f}}{\left(\phi-1\right)}=\left(\frac{n\overline{n_f}}{\phi-1}+n\right)y=n\left(y^2+y\right)$ segments, combining "two layers". This $n_2\overline{n_f}$ segment sub-chain is connected to a single seniority-2 network chain of $n$ segments, making it a combined $n_2\overline{n_f}+n=n\left(1+y+y^2\right)$ segments chain. This combining process continues sequentially, till the observed chain; there are $\phi-1$ chains of $n_\infty\overline{n_f}$ segments (Figure S.21 (c)), connected to each of the two ends of each observed chain within the material bulk; $n_\infty=n\left(1+y+y^2+y^3+\ldots\right)=n/\left(1-y\right)$. Each of these groups of $\phi-1$ chains of $n_\infty$-segments, parametrically adds a sub-chain of $\frac{n_\infty\overline{n_f}}{\phi-1}=n_\infty y$ segments to the observed chain, connecting it to the macroscopic surface, mimicking a combined chain of $n_{\mathrm{eff}}=n+2n_\infty y$ segments (Figure S.21 (d)).

$$n_{\text{eff}} = n + \frac{2ny}{1-y} = n\frac{1+y}{1-y} = n\frac{1+\dfrac{1-\Lambda}{\phi-1}}{1-\dfrac{1-\Lambda}{\phi-1}} = n\frac{\phi-\Lambda}{\phi-2+\Lambda}\ ;\ \frac{1}{n_{\text{eff}}} = \left(1-\frac{2(1-\Lambda)}{\phi-\Lambda}\right)\frac{1}{n}\ . \tag{S.44}$$

For the undeformed state ($\lambda$=1, $\Lambda = \frac{1}{n}$) with $\phi$=4, $2n_{\infty}y = n\frac{2(n-1)}{2n-1}$. For large $n$ ($\Lambda \to 0$, Gaussian-like chains), $2n_{\infty}y = n$. For any $\phi$, at $H\to 0$, $\overline{n_f}$ =1, and $2n_{\infty}y = \frac{2n}{\phi-2}$. At full extension for each chain $(\Lambda \to 1)$, the network extension is $\lambda_{\text{m}}$, and $2n_{\infty}y\big|_{\lambda\text{m}} = 0$ (eqn. (S.44)). Then the network junction fluctuations vary from being of Gaussian-like junctions when $H \to 0$, to zero fluctuations when $H \to 1$. The bridging $2n_{\infty}y$ segments from each network chain, result from mapping from junction fluctuations. The phantom network comprises $N$ effective $n$-segment network chains per unit volume, which are connected by $N_X = \left(2/\phi\right)N$ crosslink junctions per unit volume. Then, the parametric mapped $2n_{\infty}y$ segments are reverse-mapped to junction fluctuations as the effectiveness of junction fluctuations, $\eta$, which decreases as $\Lambda$ progresses. For ideal phantom networks, $\eta$ is defined with respect to a Gaussian phantom network; $\mathbf{w}_{Ph} = N_{\text{eff}}\,\mathrm{k}TW_{def}$, where $N_{\text{eff}} = (1-2\eta/\phi)N = (N-\eta N_X)$. For a large $n$ undeformed network ($\Lambda \sim 0$), $\eta = 1$. It is a finite but Gaussian-like phantom network; $\mathbf{w}_{\text{G}} = (N-N_X)\mathrm{k}TW_{def}$. With deformation, the crosslink fluctuations decrease; i.e., the network transforms from the Gaussian-like finite phantom state, and approaches the ideal affine state, $\eta = 0$.

$$\eta = \frac{\phi(1-\Lambda)}{\phi-\Lambda} \tag{S.45}$$

$$N_{\text{eff}}\,\mathrm{k}T = \frac{\rho \mathrm{R}T}{n_{\text{eff}}M_S} = \left(1-\frac{2(1-\Lambda)}{\phi-\Lambda}\right)\frac{1}{n}\frac{\rho \mathrm{R}T}{M_S} \tag{S.46}$$

For the $m$=1 model,

$$\mathbf{W}_{Ph} = \frac{1}{2n}\left(1-\frac{2(1-\Lambda)}{\phi-\Lambda}\right)\left(\left(0.5n^{0.75}+n\right)\left(\Lambda-\frac{1}{n}\right)-\left(2n-3-0.5n^{0.75}\right)\ln\left(\frac{1-\Lambda}{1-(1/n)}\right)\right) \tag{S.47}$$

$W_{def} = W_1 + W_L$ (eqn. (S.47)); $W_1 \propto \left(\Lambda - \frac{1}{n}\right)$; $W_L \propto \ln\left(\frac{1-\Lambda}{1-(1/n)}\right)$. Then, analogous to equation (S.42),

$$\sigma_{Ph} = \frac{\rho \mathrm{R}T}{M_S}\left(\frac{d\mathbf{W}_{Ph}}{d\Lambda}\right)\left(\frac{d\Lambda}{d\lambda}\right) = \frac{\rho \mathrm{R}T}{M_S}\left(\frac{d}{d\Lambda}\left(\left(1-\frac{2(1-\Lambda)}{\phi-\Lambda}\right)\frac{W_{def}}{n}\right)\right)\left(\frac{2}{3n}\left(\lambda-\frac{1}{\lambda^2}\right)\right) \tag{S.48}$$

For a phantom network of ideal FJCs, the stress components follow $W_{def} = W_1 + W_L$:

$$\mathbf{\Sigma}_{Ph} = \frac{d}{d\Lambda}\left(\left(1 - \frac{2(1-\Lambda)}{\phi - \Lambda}\right)\frac{W_{def}}{n}\right) = \mathbf{\Sigma}_{Ph,1} + \mathbf{\Sigma}_{Ph,L} \quad \text{(S.49)}$$

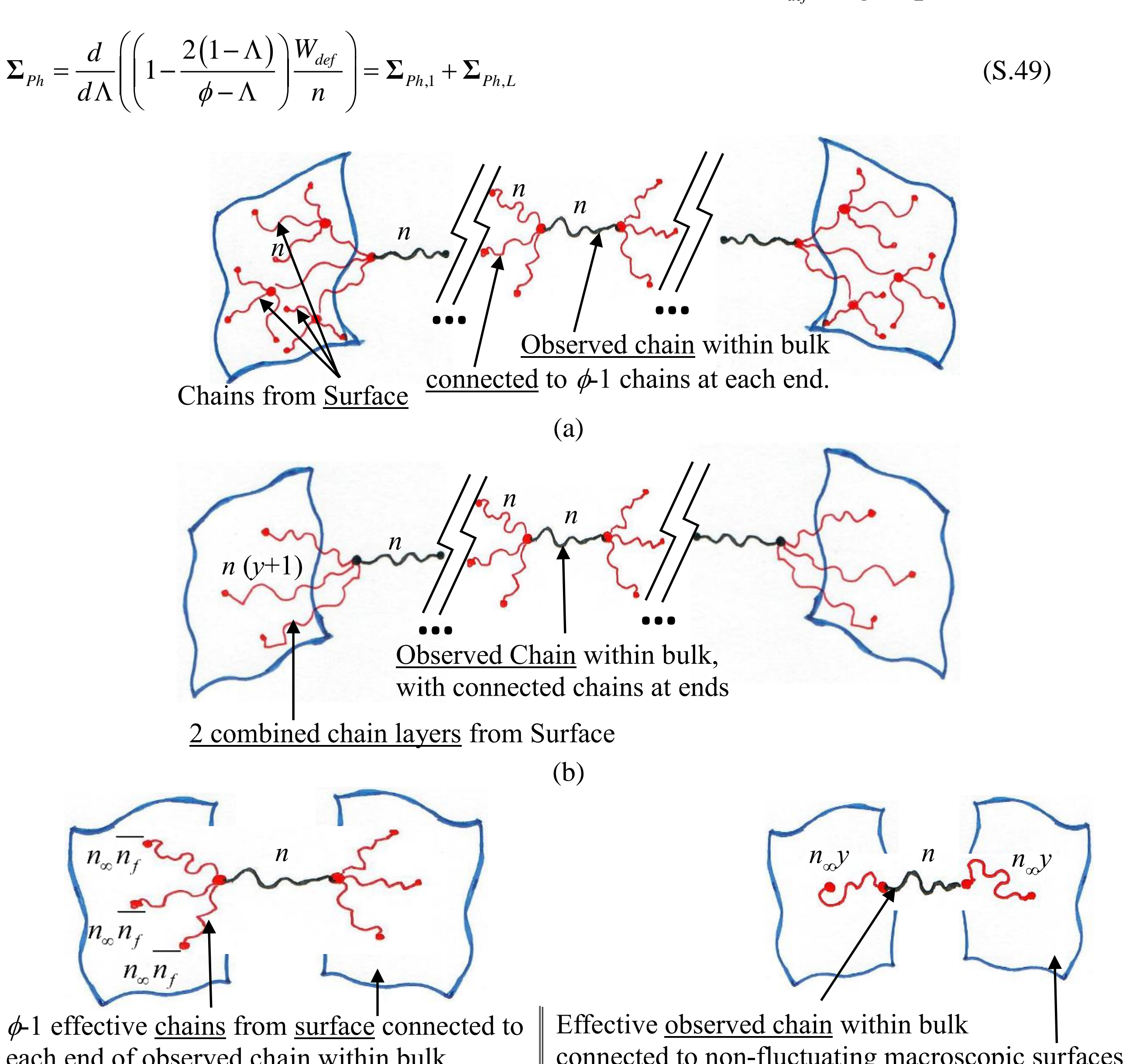

Figure S.21. (a) Tracing the path from the macroscopic surfaces to each network chain via fluctuations of junctions of connected chains. (b) At the surface, each group of $\phi-1$ zero-seniority network chains in parallel, combine in series with a seniority-1 network chain to form a combined chain of $n(y+1)$ segments. $\phi-1$ such chains connect to a seniority-2 network chain (c) $\phi-1$ chains of $n_\infty \overline{n_f}$ segments on either side of the observed chain (d) each observed chain is bridged via a connected path of fluctuating junctions to the surface, which mimics a parametric FJC of $n_\infty y$ segments from each junction end.

$$\mathbf{\Sigma}_{Ph,1} = \frac{\left(0.5n^{0.75} + n\right)}{2n}\left(\left(1 - \frac{2(1-\Lambda)}{\phi - \Lambda}\right) + \left(1 - \frac{(1-\Lambda)}{\phi - \Lambda}\right)\left(\frac{2}{\phi - \Lambda}\right)\left(\Lambda - \frac{1}{n}\right)\right) \quad \text{(S.50)}$$

$$\mathbf{\Sigma}_{Ph,L}=\frac{\left(2n-0.5n^{0.75}-3\right)}{2n}\left(\left(1-\frac{2\left(1-\Lambda\right)}{\phi-\Lambda}\right)\frac{1}{\left(1-\Lambda\right)}-\left(1-\frac{\left(1-\Lambda\right)}{\phi-\Lambda}\right)\frac{2}{\left(\phi-\Lambda\right)}\ln\left(\frac{1-\Lambda}{1-\left(1/n\right)}\right)\right) \quad \text{(S.51)}$$

Similarly, for the $m$=3 model,

$$\mathbf{W}_{Ph}=\left(1-\frac{2\eta}{\phi}\right)\frac{W_{def}}{n}=\left(1-\frac{2\left(1-\Lambda\right)}{\phi-\Lambda}\right)\frac{1}{n}\left(\begin{array}{l}\frac{1}{2}\left(1+\frac{2}{n}+n\right)\left(\Lambda-\frac{1}{n}\right)-\left(n-2\right)\ln\left[\frac{1-\Lambda}{1-\left(1/n\right)}\right]\\ -\frac{1}{20}n\left(\Lambda^{2}-\frac{1}{n^{2}}\right)-\frac{1}{15}n\left(\Lambda^{3}-\frac{1}{n^{3}}\right)\end{array}\right) \quad \text{(S.52)}$$

We can write equation (S.52) as $W_{def}=W_1+W_2+W_3+W_L$; $W_i\propto\left(\Lambda^i-\frac{1}{n^i}\right)$ and $W_L\propto\ln\left[\frac{1-\Lambda}{1-\left(1/n\right)}\right]$.

Analogous to equation (S.49),

$$\mathbf{\Sigma}_{Ph}=\frac{d\mathbf{W}_{Ph}}{d\Lambda}=\mathbf{\Sigma}_{Ph,1}+\mathbf{\Sigma}_{Ph,2}+\mathbf{\Sigma}_{Ph,3}+\mathbf{\Sigma}_{Ph,L} \quad \text{(S.53)}$$

The stress contributions correspond to the respective $W_{def}$ contributions.

$$\mathbf{\Sigma}_{Ph,1}=\frac{1}{2n}\left(1+\frac{2}{n}+n\right)\left(\left(1-\frac{2\left(1-\Lambda\right)}{\phi-\Lambda}\right)+\left(1-\frac{\left(1-\Lambda\right)}{\phi-\Lambda}\right)\left(\frac{2}{\phi-\Lambda}\right)\left(\Lambda-\frac{1}{n}\right)\right) \quad \text{(S.54)}$$

$$\mathbf{\Sigma}_{Ph,2}=\frac{-1}{10}\left(\left(1-\frac{2\left(1-\Lambda\right)}{\phi-\Lambda}\right)\Lambda+\left(1-\frac{\left(1-\Lambda\right)}{\phi-\Lambda}\right)\left(\frac{1}{\phi-\Lambda}\right)\left(\Lambda^{2}-\frac{1}{n^{2}}\right)\right) \quad \text{(S.55)}$$

$$\mathbf{\Sigma}_{Ph,3}=\frac{-1}{15}\left(\left(1-\frac{2\left(1-\Lambda\right)}{\phi-\Lambda}\right)3\Lambda^{2}+\left(1-\frac{\left(1-\Lambda\right)}{\phi-\Lambda}\right)\left(\frac{2}{\phi-\Lambda}\right)\left(\Lambda^{3}-\frac{1}{n^{3}}\right)\right) \quad \text{(S.56)}$$

$$\mathbf{\Sigma}_{Ph,L}=\left(\left(1-\frac{2\left(1-\Lambda\right)}{\phi-\Lambda}\right)\frac{\left(n-2\right)}{n\left(1-\Lambda\right)}-\left(1-\frac{\left(1-\Lambda\right)}{\phi-\Lambda}\right)\left(\frac{2\left(n-2\right)}{\phi-\Lambda}\right)\ln\left(\frac{1-\Lambda}{1-\left(1/n\right)}\right)\right) \quad \text{(S.57)}$$

We next examine this transformation from a Gaussian-like phantom undeformed network, to an affine fully stretched network. Figure S.22(a) illustrates tensile $\sigma-\lambda$ plots ($n$=20, $\phi$=4) for the finite affine network (section S.2.1) and the finite phantom network (section S.2.2.1). Figure S.22(b) contains plots of the ratios, $\sigma_{Ph}/\sigma_{Af}$ and $\mathbf{W}_{Ph}/\mathbf{W}_{Af}=1-\left(2\eta/\phi\right)$, vs any uniaxial $\Lambda$. The $m$=1 and $m$=3 models superimpose on each other. Undeformed ($\Lambda=1/n$) $\sigma_{Ph}=\sigma_{Af}=0$ and $\sigma_{Ph}/\sigma_{Af}=\mathbf{W}_{Ph}/\mathbf{W}_{Af}\sim0.52$. For large $n$ $\left(\Lambda\to0\right)$, $\mathbf{W}_{Ph}/\mathbf{W}_{Af}=\mathbf{W}_{Ph\mathrm{G}}/\mathbf{W}_{Af\mathrm{G}}\sim\sigma_{Ph\mathrm{G}}/\sigma_{Af\mathrm{G}}\sim0.5$. $\Lambda<1/n$ is unphysical, though plotted.

$\sigma_{Ph}$ and $\mathbf{W}_{Ph}$, stiffen with $\Lambda$ to reach the finite affine limit at $\Lambda \to 1$. The $\sigma-\lambda$ curves cross $\sim 0.7\lambda_{\mathrm{m}}$ and $\sigma_{Ph}/\sigma_{Af}$ reaches a maximum at Λ~0.88; i.e., over ~0.64<Λ<1, $\sigma_{Ph} > \sigma_{Af}$.

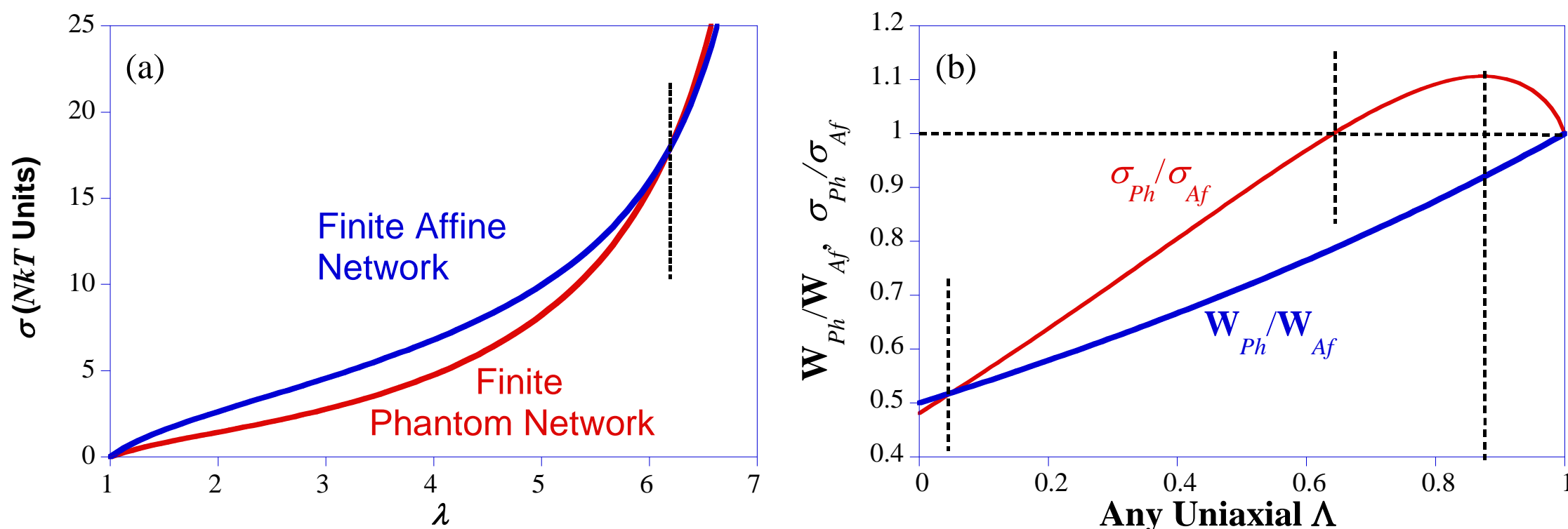

Figure S.22. Comparing Ideal Phantom and Affine Networks ($n$=20, $\phi$=4): (a) tensile $\sigma-\lambda$ plot for Phantom Networks and Affine Networks (b) $\sigma_{Ph}/\sigma_{Af}$ and $\mathbf{W}_{Ph}/\mathbf{W}_{Af}$ vs any uniaxial Λ. Phantom Network transitions to Affine Network; over ~0.64<Λ<1, $\sigma_{Ph}/\sigma_{Af}$ >1.

Simplified equations based on the Cohen's Padé approximation [7] and approximate linearization of $\mathbf{W}_{Ph}/\mathbf{W}_{Af}$, i.e., $\eta \sim (1-\Lambda)$, provide the qualitative mathematical skeleton ($m$=1 model) of ideal phantom networks; its maximum error is ~10.1% for $\phi$=3, and ~7.2% for $\phi$=4.

$\mathbf{W}_{Af} \sim \left(\Lambda - \frac{1}{n}\right) + 2\ln\left(\frac{1-\Lambda}{1-(1/n)}\right)$; then $\mathbf{\Sigma}_{Af} \sim 1 + \frac{2}{1-\Lambda}$. If $\eta \sim (1-\Lambda)$, $\frac{\mathbf{W}_{Ph}}{\mathbf{W}_{Af}} \sim \left(1 - \frac{2(1-\Lambda)}{\phi}\right)$; then

$$\mathbf{\Sigma}_{Ph} \sim \left(1 - \frac{2(1-\Lambda)}{\phi}\right)\left(1 + \frac{2}{1-\Lambda}\right) + \frac{2}{\phi}\left(\left(\Lambda - \frac{1}{n}\right) - 2\ln\left(\frac{1-\Lambda}{1-(1/n)}\right)\right).$$

### S.2.2.2 Consequences of the correct ideal phantom network

Real network models have been built on Gaussian phantom networks and on finite networks, where the surrounding constraints suppress the invariant intrinsic junction fluctuations. These constraints relax with deformation (see [18–31,33], and refs. therein). At the other extreme, chain constraints relax with deformation [35,51,54,60–63], as the network models preclude junction fluctuation. Here, refs. [31,54] consider the exact FJC expression (eqn. (S.2)) [1–3].

The ideal finite chain phantom network definition, should form the basis of further analyses of real networks. This platform could address real network behavior at two levels: (i) incorporate intrinsic chain network-level generalizations and (ii) incorporate interactions with the surrounding chains.

At the intrinsic chain network level, we can relax the assumptions for (i) uniform $n$ for all network chains or monodisperse chain molecular weight, parametrically or via simulations considering chain contour polydispersity ($n$ distribution) [42,43,50,52,64–76]. (ii) uniform segment length $l$, to a distribution of $l$, or deformable segments [43,50,77–95]. In section S.2.1.1, we have (i) addressed generalizing the undeformed chain orientation effects and (ii) suggested that the average undeformed chain vector length be monodisperse.

At the second level, the surrounding and physically entangled chains (whether Gaussian or finite) add constraints (e.g., tube and constraint models [18–31,33,35,51,54,59–63], and refs. therein). In these frameworks, the constraints from neighboring chains suppress the intrinsic phantom fluctuations of junctions (to date considered invariant Gaussian) and of the chains. These constraints are the greatest in the undeformed state, when the chains are highly coiled. With deformation, all neighbouring chains uncoil; their ability to constrain junctions and the observed chains decreases, and fluctuations increase. However, the observed chain also uncoils, and the junctions' intrinsic ability to fluctuate naturally decreases. These opposing effects need to be included while developing network elasticity models for experimental data, via an exact description of naturally deformation-dependent finite junction fluctuations and deformation-dependent surrounding constraints. The resulting $\boldsymbol{\Sigma}$, would contain both, the phantom and constraint effects. Via $\left(\rho \mathrm{R} T\right)/M_X$, it scales with $\sigma_{red} = \sigma/\left(\lambda - \left(1/\lambda^2\right)\right)$, the deformation-normalizing [96–99] Mooney-Rivlin reduced stress [4,21,26], which provides insights into the network structure when describing experimental data.

## S.3. Summary and Conclusions

The definition of an ideal phantom network evolves over three stages. (i) a closed-form framework for ideal freely jointed chains (FJCs), (ii) employed to describe quantitatively, finite fluctuations of junctions of such chains, to (iii) junction fluctuation effectiveness in the resultant elasticity relationships of ideal chain networks.

1. The first part of the framework consists of employing an accurate Padé approximation-based estimate of the inverse Langevin Function (ILF) reported by Morovati *et al.* [9], albeit with simplified coefficients. Generalizing the method of Slater *et al.* [8] (adding $1/n$ terms in the numerator), yields compact and accurate expressions that fit Treloar's exact expression [2,4] for ideal FJC distribution and accurately predicts exact ideal FJC distribution moments [2].
2. The governing geometry provides the fluctuation distribution of the junction of two ideal FJCs, as a function of chain extension. This distribution parametrically maps to an FJC with $n_{f2}$ segments.

The normalized $n_{f2}$ ($\overline{n_f} = n_{f2}\big/\left(n\!\!/_2\right)$) decreases from $\overline{n_f}$ =1.0 when the distal ends of the two FJCs coincide ($H$~0) to $\overline{n_f}$ =0, when $H$~1. If $\phi$ chains meet at a junction, the normalized $n_{f\phi}$, $\overline{n_f} = n_{f\phi}\big/\left(n/\phi\right) = 1-\Lambda$ ; $\Lambda = \left(1/(3n)\right)\left(\lambda^2 + (2/\lambda)\right)$. We describe this deformation dependence of ideal junction fluctuations, eighty years after the role of junction fluctuations was first recognized [17]; the dependence arises from the finite extensibility of ideal network FJCs, which naturally suppresses the intrinsic junction fluctuations as deformation progresses.

3. We have considered undeformed chains arranged according to BCC-type, 8-chain [41] and 4-chain network models [38,40], to ensure consistent affine, incompressible and monotonic entropic deformation. We trace the path from the non-fluctuating macroscopic surface to each observed network chain in the bulk, via the parametrically mapped deformation-dependent fluctuating junctions of the connected chains. This path reveals $\eta$, the fluctuation effectiveness of junctions, in the consequent incompressible uniaxial network elasticity relation, via the elasticity pre-factor, $N_{\text{eff}}\,\mathrm{k}T = \left(N - \eta N_X\right)\mathrm{k}T$ , which defines ideal phantom networks. The ideal phantom network is the correct platform on which to build models of surrounding constraints that impose a deformation-decreasing suppression, of naturally deformation-decreasing chain fluctuations. This combination of the phantom network with constraint models, would describe real networks.